\documentclass[aps,pre,reprint,eqsecnum,groupedaddress]{revtex4-1}

\usepackage{amsmath,amssymb}
\usepackage{bm}
\usepackage{hyperref}
\usepackage{epsfig,graphicx}

\newcommand{\ra}{\rangle}
\newcommand{\la}{\langle}

\newcommand{\be}{\begin{equation}}
\newcommand{\ee}{\end{equation}}

\newcommand{\PR}{\textit{Phys. Rev. }}
\newcommand{\PRL}{\textit{Phys. Rev. Lett.}}

\newcommand{\JSTAT}{\textit{J. Stat. Mech. (Theor. Exp.)}}

\begin{document}

\title{Nonlinear driven diffusive systems with dissipation: fluctuating hydrodynamics}

\author{A. Prados}
\email[]{prados@us.es}
\affiliation{F\'{\i}sica Te\'orica, Universidad de Sevilla, Apdo.\ de Correos 1065, Sevilla 41080, Spain}

\author{A.\ Lasanta}
\email[]{alasanta@us.es}
\affiliation{F\'{\i}sica Te\'orica, Universidad de Sevilla, Apdo.\ de Correos 1065, Sevilla 41080, Spain}

\author{Pablo I.\ Hurtado}
\email[]{phurtado@onsager.ugr.es}
\affiliation{Instituto Carlos I de F\'{\i}sica Te\'orica y Computacional, Universidad de Granada, Granada 18071, Spain}

\date{\today}

\begin{abstract}
We consider a general class of nonlinear diffusive models with bulk dissipation and boundary driving, and derive its hydrodynamic description in the large size limit. Both the average macroscopic behavior and the fluctuating properties of the hydrodynamic fields are obtained from the microscopic dynamics. This analysis yields a fluctuating balance equation for the local energy density at the mesoscopic level, characterized by two terms: (i) a diffusive term, with a current that fluctuates around its average behavior given by nonlinear Fourier's law, and (ii) a dissipation term which is a general function of the local energy density. The quasi-elasticity of microscopic dynamics, required in order to have a nontrivial competition between diffusion and dissipation in the macroscopic limit, implies a noiseless dissipation term in the balance equation, so dissipation fluctuations are enslaved to those of the density field. The microscopic complexity is thus condensed in just three transport coefficients, the diffusivity, the mobility and a new dissipation coefficient, which are explicitly calculated within a local equilibrium approximation.  Interestingly, the diffusivity and mobility coefficients obey an Einstein relation despite the fully nonequilibrium character of the problem. The general theory here presented is applied to a particular albeit broad family of systems, the simplest nonlinear dissipative variant of the so-called KMP model for heat transport. The theoretical predictions are compared to extensive numerical simulations, and an excellent agreement is found.
\end{abstract}


\maketitle

\section{Introduction}
\label{s0}

In many different contexts like biological physics, active matter, electronics, combustion, granular media, population dynamics or chemical reactions, to name just a few, the dynamics of the system of interest is characterized at the mesoscopic level by reaction-diffusion equations \cite{Fi37,bio,lutsko,MAyR11,active,PyL01,chemical,combustion,ByP96}. This type of equations often arises from the competition between diffusion and dissipation mechanisms, which typically drives the system out of equilibrium. The dissipative character of the dynamics implies a continuous loss of energy to the environment, so an steady energy input is needed in order to maintain the system in a stationary state, which is usually attained by a boundary injection mechanism. In addition, most systems in this class are strongly nonlinear, with transport coefficients which depend on the local energy density.

The physics of nonlinear driven dissipative systems is poorly understood as a result of several factors. On one hand this family of systems is intrinsically out of equilibrium: there is no equivalent to the Gibbs distribution of equilibrium systems to describe the statistics of microscopic configurations. Furthermore, there are currents of mass or energy induced by local gradients, which in turn are controlled by the dissipation and not by boundary conditions, as in standard, \emph{conservative} nonequilibrium systems. In addition, microscopic dynamics in driven dissipative media is typically irreversible, which leads to complications in their statistical description.

Despite these difficulties, there have been recent advances in nonequilibrium physics which are opening new avenues of research \cite{Bertini,Derrida}, offering tools to understand the macroscopic behavior of driven dissipative media starting from their microscopic dynamics \cite{Bodineau,PLyH11a}. The key idea which has triggered these developments has been the realization of the essential role played by macroscopic fluctuations, both in equilibrium and away from it. In fact, the study of fluctuation statistics of macroscopic observables provides an alternative way to obtain thermodynamic potentials, complementary to the usual ensemble description. This observation, valid both in equilibrium \cite{Landau} and nonequilibrium \cite{Bertini,Derrida}, is however most relevant in the latter case, where no general bottom-up approach connecting microscopic  dynamics to macroscopic properties has been found yet. The large deviation function (LDF) controlling the statistics of these fluctuations plays in nonequilibrium statistical mechanics a role similar to the equilibrium free energy \cite{Ellis,Touchette}, and the computation of LDFs in different nonequilibrium systems has thus become one of the main objectives of nonequilibrium statistical physics in recent years, triggering an enormous research effort which has led to some remarkable results \cite{Ellis,GC,LS,Bertini,Derrida,BD,Touchette,Pablo,iso}. The calculation of LDFs from first principles (i.e. from microscopic dynamics) is typically a daunting task, which has been accomplished only for a handful of models, in most cases simple interacting lattice gases \cite{Bertini,Derrida}. There is however an alternative theoretical framework, the Macroscopic Fluctuation Theory (MFT) of Bertini, De Sole, Gabrielli, Jona-Lasinio and Landim \cite{Bertini}, which studies dynamic fluctuations in diffusive media at a mesoscopic level, and offers explicit predictions for LDFs arbitrarily far from equilibrium. The starting point for this theoretical scheme is the fluctuating hydrodynamic equations describing at the mesoscopic level the evolution of the system of interest. From these equations, using a path integral formulation, one may obtain the probability of paths in mesoscopic phase space associated to a given fluctuation of a macroscopic observable, and from those probabilities a variational problem for the optimal path responsible of a given fluctuation and the associated LDF is derived \cite{Bertini,Derrida,BD,Touchette,Pablo,iso}.

The above advances have been mostly restricted to \emph{conservative} nonequilibrium systems, where dissipation is absent and nonequilibrium conditions are solely induced via boundary gradients or external fields (see however refs. \cite{Bodineau,PLyH11a}). In order to extend the ideas of Macroscopic Fluctuation Theory to nonlinear driven dissipative systems, it is of utmost importance to develop minimal models for this broad class of systems which, while capturing their essential ingredients (namely nonlinear diffusion, bulk dissipation and boundary driving), are simple enough to be amenable to both analytical calculations and extensive computer simulations. In particular, as the starting point of MFT is the mesoscopic hydrodynamic description of the system at hand, it is essential to obtain the fluctuating hydrodynamic equations and the associated transport coefficients.

In this work, we analyze the hydrodynamic behavior of a general class of systems whose main ingredients are diffusion, dissipation and boundary driving, which together with nonlinear behavior are the fingerprints of many realistic driven dissipative media. In our family of models there is one particle at each site of a $d$-dimensional lattice, and  the state of each particle is completely characterized by its ``energy''. The dynamics is stochastic and proceeds via \emph{collisions} between nearest-neighbor particles, at a rate which is a general function of the pair energy. In a collision, a certain fraction of the pair energy is dissipated, and the remaining energy is randomly distributed between the two particles. In addition, the system may be coupled to boundary thermal baths.

In the large system size limit, both continuous space and time variables can be introduced, as well as the ``hydrodynamic'' fields: energy density $\rho(x,t)$, current $j(x,t)$ and dissipation $d(x,t)$. The time evolution of the energy density follows a fluctuating balance equation of the form
\be
\partial_t\rho(x,t)=-\partial_x j(x,t)  + d(x,t) \, . \nonumber
\ee
Interestingly, the microscopic dynamics must be quasi-elastic in order to ensure that both diffusion and dissipation take place over the same time scale in the continuum limit. Using a local equilibrium approximation, the current and dissipation fields can be expressed as functions of the local energy density. In particular, the fluctuating current can be written as $j=-D(\rho)\partial_x \rho + \xi$, where the first term is nothing but Fourier's law with a diffusivity $D(\rho)$, and the second term $\xi$ is a noise perturbation, white and gaussian.  The current noise amplitude is $\sigma(\rho)/L$ (i.e., the noise strength scales as $L^{-1/2}$), where $L$ is the system size and $\sigma(\rho)$ is often referred to as the mobility in the literature. These gaussian fluctuations are expected to emerge for most situations in the appropriate mesoscopic limit as a result of a central
limit theorem. Microscopic interactions can be highly complicated, but the ensuing fluctuations of the slow hydrodynamic fields result from the sum of an enormous amount of random events at the microscale which give rise to Gaussian statistics at the mesoscale. In the present case, a proof of the gaussian character of the noise can be given, due to the simplicity of the class of models considered.  On the other hand, the dissipation field can be written as $d=-\nu R(\rho)$, where $\nu$ is a macroscopic dissipation coefficient which can be related to the inelasticity of the underlying microscopic dynamics, and $R(\rho)$ is a new transport coefficient. The dissipation field has no intrinsic noise, so its observed fluctuations
are enslaved to those of the density. This stems from the subdominant role of the noise affecting the dissipative term: its strength scales as $L^{-3/2}$ as a consequence of the quasi-elasticity of the microscopic dynamics, so it is negligible against the current noise in the mesoscopic limit.

The class of models introduced in this paper represents at a coarse-grained level the physics of many dissipative systems of technological as well as theoretical interest. In particular, when the collision rate is constant (i.e. independent of the pair energy), the Kipnis-Marchioro-Presutti (KMP) model \cite{kmp} for heat conduction is recovered  in the conservative limit. The KMP model plays a main role in nonequilibrium statistical physics as a benchmark to test theoretical advances \cite{kmp,BD,Pablo,GC,LS,iso}. For the dissipative case, different generalizations of the KMP model have been recently proposed \cite{Levine,PLyH11a}. Our general class of models contains the essential ingredients characterizing most dissipative media, namely: (i) nonlinear diffusive dynamics, (ii) bulk dissipation, and (iii) boundary injection. Moreover, it can be regarded as a  \textit{toy} model for dense granular media: particles cannot freely move but may collide with their nearest neighbors, losing a fraction of the pair energy and exchanging the rest thereof randomly. The inelasticity parameter can be thus considered as the analogue to the restitution coefficient in granular systems \cite{PyL01}.

In a forthcoming paper \cite{PLyH12a} we will use the fluctuating hydrodynamic picture that emerges from this work as starting point to analyze the large deviation function of the dissipated energy for this general class of nonlinear driven dissipative systems. In order to do so, we will extend the tools of Macroscopic Fluctuation Theory to this broad class of systems, and test our results in advanced Monte Carlo simulations capable of probing the rare events associated to the tails of the LDFs of interest.

The remainder of paper is organized as follows. In Section II we define the general class of nonlinear driven dissipative models mentioned above and discuss some limits of interest. Section III is devoted to the derivation of the hydrodynamic evolution equation for this general family of models, which is of reaction-diffusion type. In Section IV we study the fluctuating corrections to the hydrodynamic equation, deriving in this way the fluctuating hydrodynamics for our family of models, from which a full characterization of their large deviations statistics can be obtained via Macroscopic Fluctuation Theory \cite{Bertini,PLyH12a}. It is shown that both the current and dissipation noises are white and gaussian, but the dissipation noise is subdominant against the current noise in the large system size limit. In Section V we study in detail a particular family of models, for which the collision rate is a power of the local energy. We solve the full hydrodynamic problem, and compare our results with extensive numerical simulations, finding excellent agreement. Furthermore, this agreement extends also to the transport coefficients and the fluctuations of the hydrodynamic fields. A summary of the main results of the paper, together with a physical discussion thereof, is given in sec. \ref{s5}. The appendices deal with some technical details that, for the sake of clarity, we have preferred to omit in the main text.

\section{A general class of nonlinear driven dissipative models}
\label{s2}

\begin{figure}
\vspace{0.5cm}
\centerline{
\includegraphics[width=8cm]{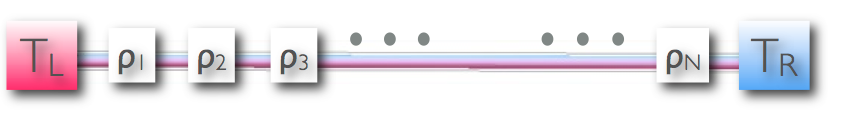}
}
\caption{\small The model is defined on lattice sites, each one characterized by an energy $\rho_l$. The dynamics is stochastic and proceeds via random collisions between nearest neighbors where part of the pair energy is dissipated to the environment and the rest is randomly redistributed within the pair. Such dynamics mimics at the mesoscopic level the evolution of a wide class of systems characterized by a nontrivial competition between diffusion and dissipation.}
\label{sketch}
\end{figure}

Let us consider a general class of models whose main ingredients are diffusion, dissipation and boundary driving. For the sake of simplicity, we will present them for the one-dimensional (1D) case, but the extension to arbitrary dimension is straightforward. The system is thus defined in a 1D lattice with $N$ sites. A configuration at a given time step $p$ is given by $\bm{\rho}=\{\rho_{l,p}\}$, $l=1,\ldots,N$, where $\rho_{l,p} \geq 0$ is the \emph{energy} of the $l$-th site at time $p$. Thus, the total energy of the system at this time is $E_p=\sum_{l=1}^N \rho_{l,p}$. The dynamics is stochastic and proceeds as follows. In an elementary step, a nearest neighbor pair of sites $(l,l+1)$ ``collides'' with probability
\begin{equation}\label{2.0}
   P_{l,p}(\bm{\rho})=\frac{f(\Sigma_{l,p})}{\sum_{l'=1}^L f(\Sigma_{l',p})}, \quad \Sigma_{l,p}=\rho_{l,p}+\rho_{l+1,p},
\end{equation}
where $f$ is a given function of the pair energy $\Sigma_{l,p}$, and $L$ is the number of possible pairs. Clearly $L\sim N$, but the particular relation depends on the boundary conditions (e.g., $L=N+1$ for open boundary conditions while $L=N$ for the periodic case). Once a pair is chosen, a certain fraction of its energy, namely  $(1-\alpha)\Sigma_{l,p}$, is dissipated to the environment. The remaining energy $\alpha \Sigma_{l,p}$ is then randomly redistributed between both sites,
\begin{equation}\label{2.1}
    \rho_{l,p+1}=z_p \alpha \Sigma_{l,p} \, , \quad \rho_{l+1,p+1}=(1-z_p) \alpha \Sigma_{l,p} \,,
\end{equation}
with $z_p$ an homogeneously distributed random number in the interval $[0,1]$.  This dynamics defines the evolution of all bulk pairs, $l=1,\ldots,N-1$. In addition, and depending on the boundary conditions imposed, boundary sites might interact with thermal baths at both ends, possibly at different temperatures $T_L$ (left) and $T_R$ (right). In this case the dynamics is
\begin{equation}\label{2.2}
    \rho_{1,p+1}=z_p \alpha (e_{1,p}+\widetilde{e}_{L}), \qquad \rho_{N,p+1}=z_p \alpha (e_{N,p}+\widetilde{e}_{R}),
\end{equation}
when the first (last) site interacts with its neighboring thermal reservoir. Here $\widetilde{e}_{\nu}$, $\nu=L,R$, is an energy randomly drawn at each step from the canonical distribution at temperature $T_{\nu}$, i.e. with probability $\text{prob}(\widetilde{e}_{\nu})=T_{\nu}^{-1} \exp (-\widetilde{e}_{\nu}/T_{\nu})$ (our unit of temperature is fixed by making $k_B=1$), see Fig. \ref{sketch}. We may consider instead an isolated system with periodic boundary conditions, such that $L=N$ and Eqs. (\ref{2.0}) and (\ref{2.1}) remain valid for $l=0$ ($l=N$) with the substitutions $\rho_{0,p}=\rho_{N,p}$ ($\rho_{N+1,p}=\rho_{1,p}$).

The simplest dynamics corresponds to  $f(\Sigma_{l,p})=1$ in Eq. (\ref{2.0}). In this case all (nearest neighbor) pairs collide with equal probability $P_{l,p}=L^{-1}$, independently of their energy. This choice (together with $\alpha=1$ above) corresponds to the Kipnis-Marchioro-Presutti (KMP) model of heat conduction \cite{kmp}, which can be considered as a coarse-grained description of  a large class of 1D diffusive systems of technological and theoretical interest. It also plays a main role in non-equilibrium statistical mechanics as a benchmark to prove rigorous results and test theoretical advances. For instance, it is one of the very few instances where Fourier's law can be rigorously proved \cite{kmp}. In addition, the KMP model has been used to investigate the validity of the additivity principle for current fluctuations \cite{BD} and the Gallavotti-Cohen fluctuation theorem \cite{GC} and its generalization in refs. \cite{Pablo,iso}. Another simple, but physically relevant, choice is $f(\Sigma_{l,p})=\Sigma_{l,p}$, so that $P_{l,p}\sim\Sigma_{l,p}/(2E_p)$ for a large system. A variant of this model has been recently used to study compact wave propagation in microscopic nonlinear diffusion \cite{HyK11}. In general, as the total exit rate from configuration $\bm{\rho}$ --the denominator of Eq. (\ref{2.0})-- is a sum of $L$ terms, we expect that $P_{l,p}\propto L^{-1}$, so it is convenient to write
\begin{equation}\label{2.b}
    P_{l,p}(\bm{\rho})=\frac{f(\Sigma_{l,p})}{L \Omega_p(L)} , \quad \text{with}\quad  \Omega_p(L)\equiv \frac{1}{L}\sum_{l=1}^L f(\Sigma_{l,p}) \, ,
\end{equation}
so $\Omega_p(L)$ remains finite as $L\to\infty$. Throughout this section, we will analyze the stochastic process generated by the dynamics (\ref{2.0})-(\ref{2.1}) supplemented with the appropriate boundary conditions.

Note that energy is conserved in the dynamics only for $\alpha=1$, while it is continuously dissipated for any $0\leq \alpha<1$. In this way, the parameter $\alpha$ can be considered as the equivalent of a restitution coefficient, using the language of granular materials. Thus, in an isolated system (without boundary driving) the energy would decrease monotonically in time. However, if energy is injected, for instance via coupling to boundary thermal baths as described above, a steady state will be eventually reached where energy injection and dissipation balance each other. In the conservative case $\alpha=1$, the transition rates introduced above verify detailed balance with the ``microcanonical'' distribution, i.e. the distribution giving equal probabilities to all the microstates compatible with the total energy of the system $E$. This implies that a closed system  would eventually approach the microcanonical equilibrium distribution for any choice of the function $f$ appearing in the transition rates. Therefore, a large subsystem, comprising a large number of sites, would be described by the canonical distribution with its temperature imposed by the rest of the system, which acts as a thermal reservoir. Thus, if we take a part of the system containing a large number of sites, it is expected on a physical basis that this region would approach the ``local equilibrium'' state compatible with the average values of the relevant macroscopic quantities, if the latter follow a very slow dynamics as compared to the relaxation of the system to the local equilibrium state. Afterwards, over the much slower ``hydrodynamic'' scale, the relevant macroscopic quantities evolve according to the macroscopic equations and reach a steady state compatible with the boundary conditions imposed on the system.

\section{Hydrodynamic Description}

The dynamics defined in Eq. (\ref{2.1}) can be rewritten as
\begin{widetext}
\begin{equation}
  \rho_{l,p+1}^{(\kappa)} = \rho_{l,p}^{(\kappa)} \left( 1-\delta_{y_p^{(\kappa)},l}-\delta_{y_p^{(\kappa)},l-1} \right)+
  z_p^{(\kappa)} \alpha \left(\rho_{l,p}^{(\kappa)}+\rho_{l+1,p}^{(\kappa)} \right) \delta_{y_p^{(\kappa)},l}+(1-z_p^{(\kappa)})\alpha \left[ \rho_{l-1,p}^{(\kappa)} +\rho_{l,p}^{(\kappa)} \right] \delta_{y_p^{(\kappa)},l-1} \label{2.3}
\end{equation}
\end{widetext}
for each realization $\kappa$ (trajectory) of the stochastic process. Here $\delta_{i,j}$ is Kronecker's delta, $z_p^{(\kappa)}$ is a random number homogeneously distributed in $[0,1]$ which controls the local energy exchange, and $y_p^{(\kappa)}\in[1,L]$ is an independent random integer which selects the colliding pair. The (conditional) probability for $y_p$ being equal to $l$, provided that the system is in a certain configuration $\bm{\rho}$, is given by $P_{l,p}(\bm{\rho})$, see Eqs.(\ref{2.0}) and (\ref{2.b}),
\begin{equation}\label{2.3b}
    \langle \delta_{y_p,l} \rangle_{\bm{\rho}} =\frac{f(\Sigma_{l,p})}{L \Omega_p(L)}\, .
\end{equation}
The energy current involved in a collision of a pair $(l,l+1)$, denoted here by $j_{l,p}$, is just the net energy traveling to the right in an unit time once dissipation takes place. Mathematically, again for each trajectory $\kappa$ of the system,
\begin{eqnarray}\label{2.4}
    j_{l,p}^{(\kappa)} & = & (\alpha \rho_{l,p}^{(\kappa)}-\rho_{l,p+1}^{(\kappa)}) \delta_{y_p^{(\kappa)},l} \nonumber \\
    & = & \alpha \left[ (1-z_p^{(\kappa)}) \rho_{l,p}^{(\kappa)}-z_p^{(\kappa)} \rho_{l+1,p}^{(\kappa)} \right] \delta_{y_p^{(\kappa)},l}.
\end{eqnarray}
where we used Eq. (\ref{2.1}) in the second equality. Equivalently, we define the dissipated energy at \emph{site} $l$ at time $p$ over each trajectory as
\begin{equation}\label{2.9}
   d_{l,p}^{(\kappa)}=-(1-\alpha)\rho_{l,p}^{(\kappa)} \left( \delta_{y_p^{(\kappa)},l}+\delta_{y_p^{(\kappa)},l-1} \right).
\end{equation}
when the colliding pair is either $(l,l+1)$ or $(l-1,l)$. Notice that, as opposed to the current term, the dissipation is defined for each site $l$ instead that for each pair $(l,l+1)$. This definition simplifies the analysis below. Making use of the current and dissipation definitions, the evolution equation (\ref{2.3}) can be rewritten in the following way,
\begin{equation}\label{2.10}
    \rho_{l,p+1}^{(\kappa)}-\rho_{l,p}^{(\kappa)}=j_{l-1,p}^{(\kappa)} -j_{l,p}^{(\kappa)} + d_{l,p}^{(\kappa)},
\end{equation}
that is a discrete reaction-diffusion equation, with a diffusive contribution $j_{l-1,p}-j_{l,p}$ and a sink term $d_{l,p}$. The latter is proportional to $1-\alpha$, thus vanishing in the elastic case $\alpha=1$.

The hydrodynamic (or average) evolution equation is obtained by summing over all the possible trajectories of the system $\kappa$, i.e. if $A$ is a certain physical property of the system, $\langle A \rangle =\sum_{\kappa=1}^{N_T} A^{(\kappa)}/N_T$, where $A^{(\kappa)}$ is the value of $A$ in the $k$th realization of the stochastic process, and the number of trajectories $N_T\to\infty$. This procedure is equivalent to averaging over all the possible sequences of the pair of independent random numbers $(z_p,y_p)$ which determine the microscopic dynamics. This said, and in order not to clutter our formulas, we will drop the superindex $\kappa$ in the remainder of the paper, as all expressions where the pair of random numbers $(z_p,y_p)$ appears are also valid for each trajectory of the system. Averaging in this way Eq. (\ref{2.10}) we arrive at
\begin{equation}\label{2.12}
    \langle \rho_{l,p+1}-\rho_{l,p}\rangle=\langle j_{l-1,p} - j_{l,p}\rangle + \langle d_{l,p}\rangle \, .
\end{equation}
Let us consider first the average value of the energy current. From Eq. (\ref{2.4}),
\begin{equation}\label{2.13}
\langle j_{l,p}\rangle=\frac{\alpha}{2L}\Big\langle  \frac{(\rho_{l,p}-\rho_{l+1,p}) f(\Sigma_{l,p})}{\Omega_p(L)} \Big\rangle,
\end{equation}
where we have taken into account Eq. (\ref{2.3b}) and $\langle z_p\rangle=1/2$. On the other hand, the average value of the dissipation is
\begin{equation}\label{2.14}
    \langle d_{l,p} \rangle=- \frac{1-\alpha}{L} \Big\langle \rho_{l,p} \left[ \frac{f(\Sigma_{l,p})+f(\Sigma_{l-1,p})}{\Omega_p(L)}\right] \Big\rangle
\end{equation}
In the large system size $L\to\infty$, we will be interested in the density, current and dissipation fields, that are smooth functions of the continuous spatial variable
\begin{equation}\label{2.15}
    x=\frac{l}{L}-\frac{1}{2},  \qquad \Delta x\equiv x_{l+1}-x_l=\frac{1}{L} ,
\end{equation}
with $x\in [-1/2,1/2]$. In this continuum limit, the average density $\langle \rho_{l,p}\rangle$ will be replaced by $\rho_{\text{av}}(x,t)$,
\begin{equation}\label{2.15b}
    \langle \rho_{l,p}\rangle \to \rho_{\text{av}}(x,t),
\end{equation}
where $t$ is a continuous time that will be introduced later on. On the other hand, because of the \textit{discrete derivative} inside the average in (\ref{2.13}), $\langle j_{l,p}\rangle$ is expected to scale as $L^{-2}$,
\begin{subequations}\label{2.16}
\begin{equation}\label{2.16a}
    \langle j_{l,p}\rangle \to L^{-2} \tau_p j_{\text{av}}(x,t),
\end{equation}
\begin{equation}\label{2.16b}
  \tau_p=\lim_{L\to\infty} \langle \Omega_p^{-1}(L)\rangle.
\end{equation}
\end{subequations}
We have introduced the term $\tau_p$ above  for the sake of convenience, because $\langle \Omega_p^{-1}(L)\rangle$ has a finite limit as $L\to\infty$. Since for each configuration of the system $\Omega_p$ is its total exit rate, $\tau_p$ is a sort of microscopic time scale, that depends on the choice of the collision rate function $f(\Sigma)$. Finally, the dissipative term $\langle d_{l,p}\rangle$ in Eq. (\ref{2.14}) scales as $(1-\alpha)L^{-1}$. From this discussion, the diffusive term in the evolution equation (\ref{2.12}) for the average density,  $\langle j_{l-1,p} - j_{l,p}\rangle$, should scale as $L^{-3}$,
\begin{equation}\label{2.17}
   \langle j_{l-1,p} - j_{l,p}\rangle \to -L^{-3} \tau_p \partial_x j_{\text{av}}(x,t)
\end{equation}
and this diffusive term should be neglected in the thermodynamic limit, unless the ``microscopic'' dissipation parameter scales as $1-\alpha \propto L^{-2}$. In fact, this is the correct scaling for the microscopic ``inelasticity'' $1-\alpha$ in the large system size limit, being the only one which guarantees that both diffusion and dissipation interplay on the same time scale at the mesoscopic level \cite{Bodineau}. With this scaling,
\begin{equation}\label{2.18}
    \langle d_{l,p}\rangle \to L^{-3} \tau_p d_{\text{av}}(x,t), \quad 1-\alpha\equiv\frac{\nu}{2L^2},
\end{equation}
where $\nu$ is a ``mesoscopic'' dissipation coefficient which remains finite in the large system size limit $L\to\infty$.

In order to be more concrete, we have to evaluate the averages on the rhs of Eqs. (\ref{2.13}) and (\ref{2.14}). This can be done within a local equilibrium approximation (see appendix \ref{apa}), which we expect to be valid in the thermodynamic limit  \cite{Bertini,Sp9091}. For large system sizes, a clear time scale separation appears between the scale (of the order of several times $\tau_p$) in which the system approaches locally the equilibrium distribution (i.e. a local product Gibbs measure with local \emph{temperature} $\langle\rho_{l,p}\rangle$) and the much longer time scale over which the average density field evolves following the hydrodynamic equation. Under this local equilibrium approximation, the average current verifies
\begin{equation}\label{2.19}
    \langle j_{l,p}\rangle \sim -L^{-2} \tau_p D(\langle\rho_{l,p}\rangle) \frac{\langle \rho_{l+1,p}-\rho_{l,p}\rangle}{\Delta x},
\end{equation}
with the diffusivity $D(\rho)$ given by
\begin{equation}\label{2.20}
    D(\rho)=\frac{1}{6}\int_0^\infty dr\, r^7 f(\rho r^2) e^{-r^2}.
\end{equation}
Equation (\ref{2.19}) is thus consistent with the scaling in Eq. (\ref{2.16a}). Furthermore,
\begin{equation}\label{2.25}
    j_{\text{av}}(x,t)=-D(\rho_{\text{av}})\partial_x \rho_{\text{av}},
\end{equation}
i.e. Fourier's law holds with diffusivity $D(\rho_{\text{av}})$. On the other hand, the average dissipation, Eq. (\ref{2.17}), can be written within the local equilibrium approximation as
\begin{equation}\label{2.21}
    \langle d_{l,p}\rangle \sim -L^{-3}  \tau_p \nu R(\langle \rho_{l,p}\rangle),
\end{equation}
with
\begin{equation}\label{2.22}
    R(\rho)=\rho \int_0^\infty dr r^5 f(\rho r^2) e^{-r^2}.
\end{equation}
Again, Eq. (\ref{2.21}) is consistent with the previously assumed scaling (\ref{2.18}), with
\begin{equation}\label{2.26}
    d_{\text{av}}(x,t)=-\nu R(\rho_{\text{av}})
\end{equation}
Interestingly, this new transport coefficient $R(\rho)$, associated to the dissipation, can be related to the diffusivity. By differentiating Eq. (\ref{2.22}) with respect to $\rho$ after a change of variables $z=r \sqrt{\rho}$, it is found that
\begin{equation}\label{2.26b}
  D(\rho)=\frac{1}{6}\frac{dR(\rho)}{d\rho}+\frac{R(\rho)}{3\rho}.
\end{equation}
In fact, given the diffusivity, this equation can be considered as a first order differential equation for $R(\rho)$. The solution thereof, with an appropriate boundary condition (normally $R(\rho=0)=0$), is equivalent to calculate the integral in Eq. (\ref{2.22}).

Now, by introducing a macroscopic time $t$, such that its time increment at step $p$ is given by
\begin{equation}\label{2.23}
   \Delta t_p=L^{-3} \tau_p , \qquad t=\sum_{j=0}^{p-1} \Delta t_j=L^{-3} \sum_{j=0}^{p-1} \tau_j,
\end{equation}
we can rewrite the evolution equation for the average density (\ref{2.12}) as
\begin{equation}\label{2.24}
    \partial_t \rho_{\text{av}}(x,t)=-\partial_x j_{\text{av}}(x,t)+d_{\text{av}}(x,t),
\end{equation}
where the average current $j_{\text{av}}(x,t)$ verifies Fourier's law with diffusivity $D(\rho_{\text{av}})$, Eqs. (\ref{2.20})-(\ref{2.25}), and the average dissipation $d_{\text{av}}(x,t)$ is given in terms of the density field by Eqs. (\ref{2.22})-(\ref{2.26}).

The boundary conditions for eq (\ref{2.24}) depend on the physical situation of interest. For instance, we may consider that the system is kept in contact with two thermal reservoirs at $x=\pm 1/2$, at the same temperature $T$.  In that case, the system eventually reaches a steady state in the long time limit, for which the injection of energy through the boundaries and the dissipation balance each other. The stationary average (macroscopic) solution of (\ref{2.24}) verifies
\begin{equation}\label{2.27}
    j_{\text{av}}'(x)+\nu R(\rho_{\text{av}}(x))=0, \quad j_{\text{av}}(x)=-D(\rho_{\text{av}}(x)) \rho_{\text{av}}'(x),
\end{equation}
where the prime indicates spatial derivative. The first equation in (\ref{2.27}) follows from (\ref{2.24}), while the second one is Fourier's law. Equivalently, a closed second-order equation for $\rho$ may be written,
\begin{equation}\label{2.28}
    \frac{d}{dx} \left[ D(\rho_{\text{av}}) \rho'_{\text{av}} \right]=\nu R(\rho_{\text{av}}),
\end{equation}
with the boundary conditions $\rho_{\text{av}}(\pm 1/2)=T$. We may now introduce an auxiliary field $y$ that will be helpful later on,
\begin{subequations}\label{2.29}
\begin{equation}\label{2.29a}
    y=R(\rho),
\end{equation}
such that
\begin{equation}\label{2.29b}
    d(x,t)=-\nu y(x,t),
\end{equation}
\end{subequations}
Equations (\ref{2.27}) and (\ref{2.28}) can be also rewritten for the variable $y$,
\begin{equation}\label{2.30}
    j'_{\text{av}}(x)+\nu y_{\text{av}}(x)=0, \quad j_{\text{av}}(x)=-\hat{D}(y_{\text{av}}(x)) y_{\text{av}}'(x),
\end{equation}
with
\begin{equation}\label{2.31}
    \hat{D}(y)=\left(\frac{dy}{d\rho}\right)^{-1}D(\rho)=
    \left(\frac{dR(\rho)}{d\rho}\right)^{-1}D(\rho)\, .
\end{equation}
Thus, $\hat{D}$ acts as an ``effective'' diffusivity, i.e. the factor multiplying the spatial gradient when writing Fourier's equation in terms of the new variable $y$. Taking into account Eq. (\ref{2.26b}),
\begin{equation}\label{2.33}
  \hat{D}(y)=\frac{1}{6}+\frac{1}{3} \frac{d\ln \rho(y)}{d\ln y}.
\end{equation}
Interestingly, the above equation shows that $\hat{D}$ is constant, independent of $y$, whenever $y=R(\rho)$ depends algebraically on $\rho$. This observation will be very useful in Section \ref{s4}, where we study in detail a particular family of models. Equation (\ref{2.30}) can also be summarized in a second order differential equation for $y_{\text{av}}$,
\begin{equation}\label{2.34}
    \left[ \hat{D}(y_{\text{av}}) y'_{\text{av}} \right]'=\nu y_{\text{av}}, \quad y_{\text{av}}(\pm 1/2)=R(T),
\end{equation}
the solution of which gives the average density field in the situation at hand. When $\hat{D}$ is constant, Equation (\ref{2.34}) is linear in $y$ and can be readily integrated.

\section{Fluctuations of the current and dissipation fields}
\label{s3}

The local energy $\rho_{l,p}$ obeys the balance equation (\ref{2.10}), with a diffusive term, given by the discrete spatial derivative of the microscopic current, and a dissipative term. Taking averages over all possible realizations of the dynamics, the hydrodynamic equation (\ref{2.24}) has been obtained, with a similar structure but now in terms of the the continuum hydrodynamic fields (density, current and dissipation). The main objective of this section is to analyze the local fluctuations of the current and the dissipation fields away from their averages or, in other words, the properties of their respective noise terms.

We will address the problem of characterizing the current and dissipation noise terms in the limit of large system size $L \gg 1$, the same one in which Eq. (\ref{2.24}) has been shown to be valid over the hydrodynamic scale defined in Eq. (\ref{2.23}). The idea is to split the microscopic current $j_{l,p}$ into a ``main'' term $\widetilde{j}_{l,p}$, whose average coincides with $\langle j_{l,p}\rangle$, see Eq. (\ref{2.13}), and a noise term $\xi_{l,p}$. Similarly, we will have $d_{l,p}=\widetilde{d}_{l,p}+\eta_{l,p}$, with the main term $\widetilde{d}_{l,p}$ verifying $\langle \widetilde{d}_{l,p}\rangle=\langle d_{l,p}\rangle$, see Eq. (\ref{2.14}). Thus, $\eta_{l,p}$ is the dissipation noise. We will investigate the properties of both noise terms and find that, in the large system size limit: (i) both noises are white and Gaussian, and (ii) the dissipation noise is subdominant against the current noise if $1-\alpha={\cal O}(L^{-2})$, as given by Eq. (\ref{2.18}). Therefore, the fluctuations of the dissipation field are enslaved to those of the density field, and the current noise is the main source for the fluctuations in the system at the mesoscopic level.

\subsection{Fluctuating current. Average value and noise properties}
\label{s3a}

Let us consider the fluctuations of the microscopic current $j_{l,p}$. As we have already stated, the idea is to split the current into a ``main'' and a ``noise'' term, 
\begin{equation}\label{3.0}
    j_{l,p}=\widetilde{j}_{l,p}+\xi_{l,p}
\end{equation}
such that $\langle \widetilde{j}_{l,p}\rangle=\langle j_{l,p}\rangle$. It is clear from Eq. (\ref{2.13}) that the following choice
\begin{equation}\label{3.1}
    \widetilde{j}_{l,p}=\frac{\alpha}{2L} \frac{ \left( \rho_{l,p}-\rho_{l+1,p} \right) f(\Sigma_{l,p})}{\Omega_p(L)} \, ,
\end{equation}
guarantees the previous condition on the average. To stress out the difference between $j_{l,p}$ and $\widetilde{j}_{l,p}$, notice that while the former is exactly zero unless pair $(l,l+1)$ actually collides at time $p$, see Eq. (\ref{2.4}), the latter may take a nontrivial value even in the absence of such collision; however their averages coincide \cite{foot1}. Eq. (\ref{3.1}) is Fourier's law at the microscopic level, i.e. $\widetilde{j}_{l,p}$ is proportional to the local density gradient. We want to study now the properties of the noise $\xi_{l,p}=j_{l,p}-\widetilde{j}_{l,p}$, namely its average and its correlation function $\langle \xi_{l,p}\,\xi_{l',p'}\rangle$. Since $\langle \widetilde{j}_{l,p}\rangle=\langle j_{l,p}\rangle$, it is clear that
\begin{equation}\label{3.2}
    \langle \xi_{l,p}\rangle =0.
\end{equation}
For the noise correlation function, it is straightforward to show that
\begin{equation}\label{3.4}
    \langle \xi_{l,p}\, \xi_{l',p'}\rangle=\langle j_{l,p}\,j_{l',p'}\rangle - \langle \widetilde{j}_{l,p}\,\widetilde{j}_{l',p'}\rangle.
\end{equation}
Furthermore, from the current definition Eq. (\ref{2.4}), it is also easily proven that $\langle \xi_{l,p}\, \xi_{l',p'}\rangle=0$ for $p\neq p'$, so the current noise at different times is uncorrelated. For equal times, $p=p'$, the second term on the rhs of (\ref{3.4}) is negligible against the first one in the limit $L\gg 1$, because it is $\sim{\cal O}(L^{-2})$, while the leading behavior of the first term will be shown to be $\sim{\cal O}(L^{-1})$. Using now the definition of the microscopic current (\ref{2.4}),
\begin{eqnarray}
    j_{l,p} j_{l',p}&=& \alpha^2 \left[ (1-z_p) \rho_{l,p}-z_p \rho_{l+1,p} \right] \label{3.5} \\ && \times \left[ (1-z_{p}) \rho_{l',p}-z_{p} \rho_{l'+1,p} \right] \delta_{y_p,l}\delta_{y_{p},l'}.\nonumber
\end{eqnarray}
Taking into account that $\langle z_p\rangle=1/2$, $\langle z_p^2\rangle=1/3$, and
\begin{equation}\label{3.6}
   \langle \delta_{y_p,l}\delta_{y_p,l'}\rangle_{\bm{\rho}} =\frac{f(\Sigma_{l,p})}{L\Omega_p(L)} \delta_{l,l'} \, ,
\end{equation}
we arrive at
\begin{equation}\label{3.7}
     \langle j_{l,p} j_{l',p} \rangle \sim \frac{\alpha^2}{3L} \Bigg\langle \frac{\left(\rho_{l+1,p}^2+\rho_{l,p}^2-\rho_{l,p}\rho_{l+1,p}\right)
     f(\Sigma_{l,p})}{\Omega_p(L)}\Bigg\rangle \delta_{l,l'},
\end{equation}
Hence, the leading behavior of the current noise correlation in the $L\gg 1$ limit is
\begin{equation}\label{3.8}
    \langle \xi_{l,p}\, \xi_{l',p'}\rangle\sim \frac{1}{L} \Xi_{l,p} \delta_{l,l'} \delta_{p,p'},
\end{equation}
where we have neglected ${\cal O}(L^{-2})$ terms, and defined
\begin{equation}\label{3.9}
    \Xi_{l,p}= \frac{\alpha^2}{3}\Bigg\langle \frac{\left(\rho_{l+1,p}^2+\rho_{l,p}^2-\rho_{l,p}\rho_{l+1,p}\right)
    f(\Sigma_{l,p})}{\Omega_p(L)} \Bigg\rangle.
\end{equation}
Equation (\ref{3.9}) cannot be expressed in terms of $\langle\rho_{l,p}\rangle$, so the fluctuating balance equation (\ref{2.10}) is not closed at the microscopic level, as otherwise expected. Using here again the local equilibrium approximation previously employed \cite{Bertini} (see appendix \ref{apa}), we obtain
\begin{equation}\label{3.10}
    \Xi_{l,p} \sim \tau_p \frac{\alpha^2\rho_{\text{av}}^2}{3}\int_0^\infty dr r^7 f(\rho_{\text{av}} r^2) e^{-r^2}.
\end{equation}
A rigorous proof of this local equilibrium approximation can be found in \cite{Sp9091}, in the context of conservative interacting particle systems. The application thereof to the dissipative case studied here relies on the fact that the microscopic dynamics is quasielastic. Moreover, the validity of this local equilibrium approximation will be fully confirmed \emph{a posteriori} in extensive numerical simulations, see Section \ref{s4}.

In the large system size limit we have to take into account Eq. (\ref{2.16a}), and introduce a similar scaling for the fluctuating current and its noise,
\begin{equation}\label{3.11}
    j_{l,p} \to L^{-2} \tau_p j(x,t), \quad \xi_{l,p} \to L^{-2} \tau_p \xi(x,t),
\end{equation}
so that
\begin{equation}\label{3.12}
    j(x,t)=\widetilde{j}(x,t)+\xi (x,t).
\end{equation}
With this definition,
\begin{equation}\label{3.13}
    \langle \widetilde{j}(x,t)\rangle=j_{\text{av}}(x,t),
\end{equation}
and thus $\langle \xi(x,t)\rangle=0$. Besides, the correlation function of the current noise is, by combining Eqs. (\ref{3.8})-(\ref{3.11}),
\begin{equation}\label{3.14}
    \langle \xi(x,t) \xi(x',t') \rangle\sim \frac{1}{L} \sigma(\rho_{\text{av}}) \delta(x-x') \delta(t-t'),
\end{equation}
where we have taken into account the quasielasticity of the microscopic dynamics, $1-\alpha={\cal O}(L^{-2})$, as given by Eq. (\ref{2.18}). We have also introduced the so-called mobility
\begin{equation}\label{3.15}
    \sigma(\rho)=\frac{\rho^2}{3} \int_0^\infty dr \, r^7 f(\rho  r^2) e^{-r^2}=2\rho^2 D(\rho)
\end{equation}
and made use of
\begin{equation}\label{3.16}
     \frac{\delta_{l,l'}}{\Delta x}\sim\delta(x-x'), \qquad  \frac{\delta_{p,p'}}{\Delta t_p}\sim\delta(t-t') ,
\end{equation}
where $\Delta x$ and $\Delta t_p$ were defined in Eqs.(\ref{2.15}) and (\ref{2.23}), respectively. The current noise is hence white, its average value vanishes and it is delta-correlated in space. Furthermore, it can be shown that it is gaussian, and we present a sketch of the proof in appendix \ref{apb}. Remarkably, Eq. (\ref{3.15}), which relates mobility and diffusivity, $\sigma(\rho)=2\rho^2 D(\rho)$, is a fluctuation-dissipation relation for the dissipative case. In fact, it is the same one as in the conservative case, because of the quasi-elasticity of the underlying microscopic (stochastic) dynamics \cite{foot2}.

The gaussian character of the current field fluctuations will allow us, in a forthcoming paper \cite{PLyH12a}, to extend the recently-introduced Macroscopic Fluctuation Theory \cite{Bertini} to the case of nonlinear driven dissipative systems \cite{PLyH11a}. This theoretical scheme allows to study the fluctuations (both typical and rare) of macroscopic observables arbitrarily far from equilibrium, offering predictions for the large-deviation functions which characterize their statistics and the optimal paths in phase space responsible of these fluctuations. The validity of the fluctuating hydrodynamic equation derived in this paper, and in particular of the local equilibrium approximation used here, can be checked \textit{a posteriori} by comparing the predictions for the large-deviation functions that arise from the macroscopic fluctuation theory associated to our equation with the numerical results.

\vspace*{1ex}

\subsection{Fluctuating dissipation. Average value and noise properties}
\label{s3b}

Our starting point is Eq. (\ref{2.9}) for the microscopic dissipation $d_{l,p}$, that we split into a main term plus a noise term, i.e.
\begin{subequations}\label{3.17}
\begin{equation}\label{3.17a}
d_{l,p}=\widetilde{d}_{l,p}+\eta_{l,p}
\end{equation}
\begin{equation}\label{3.17b}
\widetilde{d}_{l,p}=-\frac{1-\alpha}{L} \frac{\rho_{l,p}\left[f(\Sigma_{l,p})+f(\Sigma_{l-1,p})\right]}{\Omega_p(L)}\, .
\end{equation}
\end{subequations}
Clearly the average of the main term verifies $\langle \widetilde{d}_{l,p}\rangle=\langle d_{l,p}\rangle$, see Eq. (\ref{2.14}), and thus the noise average vanishes, $\langle\eta_{l,p}\rangle=0$. From Eqs. (\ref{2.9}) and (\ref{3.17}), we have
\begin{equation}\label{3.18}
    \eta_{l,p}=-(1-\alpha)\rho_{l,p} \left[ \delta_{y_p,l}+\delta_{y_p,l-1}-\frac{f(\Sigma_{l,p})+
    f(\Sigma_{l-1,p})}{L\Omega_p(L)}\right]
\end{equation}
and
\begin{widetext}
\begin{equation}\label{3.19}
     \eta_{l,p} \eta_{l',p'}  =  (1-\alpha)^2 \rho_{l,p} \rho_{l',p'} \left[ \delta_{y_p,l}+\delta_{y_p,l-1}-
     \frac{f(\Sigma_{l,p})+f(\Sigma_{l-1,p})}{L\Omega_p(L)} \right] \left[ \delta_{y_{p'},l'}+\delta_{y_{p'},l'-1}-
     \frac{f(\Sigma_{l',p'})+f(\Sigma_{l'-1,p'})}{L\Omega_{p'}(L)}\right]
\end{equation}
Now, taking averages it is clear that $\langle \eta_{l,p}\eta_{l',p'}\rangle=0$ if $p\neq p'$, because the variables $y_p$, $y_{p'}$ are independent in that case and the terms in brackets have zero average. Then, we restrict ourselves to the case $p=p'$, for which a simple calculation similar to the one carried out for the current leads to
\begin{equation}\label{3.20}
    \langle \eta_{l,p} \eta_{l',p}\rangle\sim \frac{(1-\alpha)^2}{L} \Big\langle \frac{\rho_{l,p} \rho_{l',p}}{\Omega_p(L)}  \left[  f(\Sigma_{l,p}) (\delta_{l,l'}+\delta_{l,l'-1})+ f(\Sigma_{l-1,p}) (\delta_{l,l'}+ \delta_{l,l'+1}) \right] \Big\rangle,
\end{equation}
\end{widetext}
after having made use of Eq. (\ref{3.6}) and neglected ${\cal O}(L^{-2})$ terms. In order to give an explicit expression for this correlation, we should have to evaluate the averages in (\ref{3.20}), for instance by making use of the local equilibrium approximation, as in the previous subsection. Nevertheless, for the work presented here, this is not necessary due to the quasi-elasticity of the microscopic dynamics, as given by Eq. (\ref{2.18}). This will allow us to show that the dissipation noise is subdominant against the current noise in the large system size limit, without having to calculate explicitly its amplitude. The structure of (\ref{3.20}) implies that
\begin{eqnarray}
    \langle \eta_{l,p}\eta_{l',p'}\rangle &\sim & L^{-5} \nu^2 \tau_p \nonumber\\
   && \times \left( \kappa^{(1)}_{l,p} \delta_{l,l'}+\kappa^{(2)}_{l,p}\delta_{l,l'-1}+\kappa^{(3)}_{l,p} \delta_{l,l'+1}
    \right) \delta_{p,p'} \nonumber \\
    && \label{3.21}
\end{eqnarray}
where $\kappa^{(i)}_{l,p}$, $i=1,2,3$, are certain averages which remain of the order of unity in the limit $L\to\infty$ (they become functions of $\langle\rho_{l,p}\rangle$ in the local equilibrium approximation). With the consistent definition $\eta(x,t)=L^3 \eta_{l,p}/\tau_p$, see Eq. (\ref{2.18}), we introduce the continuum limit of the dissipation noise and its correlation function becomes
\begin{equation}\label{3.22}
    \langle \eta(x,t) \eta(x',t')\rangle\sim L^{-3} \nu^2 \kappa(\rho)  \delta(x-x')\delta(t-t'),
\end{equation}
where we have used Eq. (\ref{3.16}), and defined $\kappa(\rho)$ as the continuum limit of $\sum_{i=1}^3 \kappa_{l,p}^{(i)}$ above. In particular, it can be shown that the continuum limit of the diagonal term $\kappa_{l,p}^{(1)}$ in Eq. (\ref{3.21}) is exactly equal to $\sigma(\rho)$, while each non-diagonal term $\kappa_{l,p}^{(2,3)}$ contributes $\sigma(\rho)/4$ to $\kappa(\rho)$. Equation (\ref{3.22}) tells us that the dissipation noise ($\propto L^{-3/2}$) is much weaker than the current noise ($\propto L^{-1/2}$) in the large system size limit $L\gg 1$. Therefore, we will neglect it in the following, i.e. we consider that
\begin{equation}\label{3.23}
    d(x,t)=-\nu R(\rho),
\end{equation}
i.e. only the fluctuations affecting the current term will play a relevant role in the mesoscopic limit. We expect that this approximation becomes exact in the thermodynamic limit as $L\to\infty$. Then, the fluctuations of the dissipation $d(x,t)$ are ``enslaved'' to those of the energy field $\rho(x,t)$, as a consequence of the quasi-elasticity of the dynamics at the microscopic level. This quasi-elastic character of the microscopic dynamics is compatible with the existence of a finite dissipation at the mesoscopic level, as expressed by the finite value of the macroscopic dissipation coefficient $\nu$.

\section{A nonlinear dissipative version of the KMP model of heat transport}
\label{s4}

We now will apply the theory developed in the previous sections to a broad class of dissipative models. More concretely, we will restrict ourselves to the following choice for the collision rate function
\begin{equation}\label{4.0}
  f(\rho)=\frac{2}{\Gamma(\beta+3)}\rho^\beta,
\end{equation}
i.e., $f(\rho)\propto\rho^\beta$, with $\beta>-3$ but otherwise arbitrary.  We have introduced the constant $2/\Gamma(\beta+3)$ \cite{gamma} for the sake of convenience, as it simplifies the expressions of the transport coefficients, see below. For $\beta=0$, $f(\rho)=1$ and all the pairs collide with equal probability, independently of their energy value. Thus, the dissipative generalization of the KMP model introduced in ref. \cite{PLyH11a} is recovered. For $\beta=1$, $f(\rho)=\rho/3$ and the colliding pairs are chosen with probability proportional to their energy. The conservative case has been recently analyzed in \cite{HyK11}.

The transport coefficients for this family of models are easily calculated. The coefficient linked to the dissipation, $R(\rho)$, is readily obtained from Eq. (\ref{2.22}),
\begin{equation}\label{4.1}
    R(\rho)=\frac{2}{\Gamma(\beta+3)}\rho^{\beta+1}\int_0^\infty dr \, r^{5+2\beta} e^{-r^2}=\rho^{\beta+1},
\end{equation}
that gives the rationale behind the choice of the proportionality constant in Eq. (\ref{4.0}). The diffusivity is calculated by substituting Eq. (\ref{4.3}) into (\ref{2.26b}),
\begin{equation}\label{4.2}
    D(\rho)=\frac{\beta+3}{6}\rho^\beta.
\end{equation}
Of course, the same result is obtained with Eq.(\ref{2.20}). Finally, the mobility $\sigma(\rho)$ follows from the fluctuation-dissipation relation (\ref{3.15}), which gives it in terms of the diffusivity,
\begin{equation}\label{4.3}
    \sigma(\rho)=2\rho^2 D(\rho)=\frac{\beta+3}{3} \rho^{\beta+2}.
\end{equation}
Of course, for $\beta=0$ the values of the transport coefficients of the dissipative version of the KMP model are recovered, $D(\rho)=1/2$, $\sigma(\rho)=\rho^2$, and $R(\rho)=\rho$ \cite{PLyH11a}. Interestingly, this kind of dependence with the energy density $\rho$ appears in real systems. For instance, in granular materials \cite{PyL01} the density field $\rho$ may be assimilated to the local granular temperature. Moreover, for the hard sphere model, the average collision rate is proportional to the square root of the granular temperature. Thus, this granular gas case should correspond to $\beta=1/2$, and in fact it is found that $D(\rho)\propto\rho^{1/2}$ while the dissipative term goes as $R(\rho)\propto\rho^{3/2}$. The latter is the responsible for the algebraic decay of the granular temperature (Haff's law; $\propto t^{-2}$ for large times) observed in the homogeneous case when the system is isolated \cite{PyL01} .

For the class of models at hand, we thus have a fluctuating hydrodynamic equation
\begin{equation}\label{4.4}
    \partial_t \rho(x,t)=-\partial_x j(x,t)-\nu R(\rho(x,t)),
\end{equation}
where $j(x,t)$ is the fluctuating current,
\begin{equation}\label{4.5}
    j(x,t)=Q[\rho(x,t)]+\xi(x,t),
\end{equation}
with local average behavior given by Fourier's law,
\begin{equation}\label{4.6}
    Q[\rho]=-D(\rho)\partial_x \rho(x,t),
\end{equation}
and $\nu$ is the macroscopic dissipation coefficient defined in Eq. (\ref{2.18}). The first term in the rhs of Eq. (\ref{4.4}) accounts for the diffusive spreading of the energy, and it is also present in the conservative case, while the second one gives the rate of dissipation of energy in the bulk. The current noise term $\xi(x,t)$ describes the random fluctuations at the mesoscopic level, as has been analyzed in the previous section. In the large system size limit $L\gg 1$, it is gaussian and white with
\begin{subequations}\label{4.7}
\begin{equation}\label{4.7a}
    \langle \xi(x,t)\rangle=0, \quad
\end{equation}
\begin{equation}\label{4.7b}
    \langle \xi(x,t)\xi(x',t')\rangle=\frac{1}{L} \sigma(\rho) \delta(x-x') \delta(t-t').
\end{equation}
\end{subequations}
This gaussian fluctuating field emerges in the mesoscopic limit as a result of the central limit theorem: although microscopic interactions are rather complicated, the ensuing fluctuations of the hydrodynamic fields, which evolve over a much slower time scale, result from the sum of an enormous amount of random events at the microscale and give rise to gaussian statistics of ${\cal O}(L^{-1/2})$ at the mesoscale.

\begin{figure}
\vspace{-0.5cm}
\centerline{
\includegraphics[width=9.5cm]{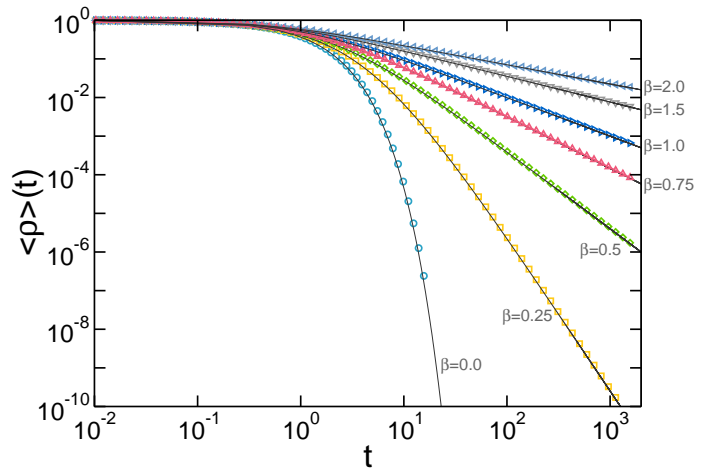}
}
\caption{\small Decay with time of the average energy per site in the isolated nonlinear dissipative KMP model. Here we plot data for $\nu=1$, $L=80$, and different values of $\beta\in[0,2]$, though qualitatively similar behavior is obtained for other values of $\nu$ and $L$. Lines are theoretical predictions, see Eq. (\ref{haffeq}).}
\label{haff}
\end{figure}

Let us investigate now the average behavior for the family of models defined by the collision rate choice (\ref{4.0}), whose transport coefficients are given by Eqs. (\ref{4.1})-(\ref{4.3}). First, we focus on the time evolution of an isolated system (i.e. with periodic boundary conditions). Due to the dissipation, and assuming no spatial structure, the system enters into an homogeneous cooling state (HCS) characterized by a continuous energy dissipation. Homogeneity implies that $\rho(x,t)=\rho(t)$, and this energy density obeys $\partial_t\rho=-\nu\rho^{\beta+1}$ according to Eqs. (\ref{4.4}) and (\ref{4.1}). The solution is
\be
\label{haffeq}
\rho(t)=
\begin{cases}
\frac{\displaystyle \rho_0}{(\displaystyle 1+\nu\beta\rho_0^\beta t)^{1/\beta}} & \beta \neq 0 \\
\phantom{aaa} & \phantom{aaa} \\
\rho_0\text{e}^{-\nu t} & \beta=0
\end{cases}
\ee
where $\rho_0$ is the initial energy density in the system (which is fixed). Notice the exponential decay of energy in time for the linear ($\beta=0$) case, while for all nonlinear cases $\beta\neq 0$ a power-law decay is obtained. This is in fact a generalization of the well-known Haff's law of granular gases, which corresponds to $\beta=1/2$ in our model. Fig. \ref{haff} shows the time evolution of the average energy density for different $\beta$'s and particular values of $\nu$ and $L$ as measured in Monte Carlo simulations. The agreement between numerical results and Eq. (\ref{haffeq}) is excellent in all cases, supporting the homogeneity conjecture used above.
We do not expect the HCS to become unstable as time proceeds because our model does not contain velocity variables whose correlation upon collision is at the origin of the well-known breakdown of HCS in $d$-dimensional ($d>1$) granular gases \cite{PyL01}.

Next, we study the steady-state behavior when the system is coupled to boundary thermal baths at equal temperature $T$. Following the procedure introduced at the end of Section \ref{s2}, it is convenient to define an auxiliary variable
\begin{equation}\label{4.8}
    y=\rho^{\beta+1}, \quad \rho=y^{\frac{1}{\beta+1}},
\end{equation}
where we have made use of Eqs. (\ref{2.29a}) and (\ref{4.1}). We also calculate the ``effective'' diffusivity $\hat{D}(y)$ defined in Eq. (\ref{2.31}),
\begin{equation}\label{4.11}
    \hat{D}=\frac{\beta+3}{6(\beta+1)},
\end{equation}
where we have made use of Eq. (\ref{2.33}). Thus, while the ``true'' diffusivity $D(\rho)$ depends on $\rho$, as given by Eq. (\ref{4.1}), the ``effective'' diffusivity  is constant, $\hat{D}(y)=\hat{D}$. This allows us to calculate explicitly the average profiles for the density and the current, since Eq. (\ref{2.34}), that determines them, is linear for $y$. The steady average solution in this boundary-driven case is thus
\begin{subequations}\label{4.12}
\begin{equation}\label{4.12a}
    y_{\text{av}}(x)=T^{\beta+1}\, \frac{\cosh\left(x\sqrt{\frac{\nu}{\hat{D}}}\right)}
    {\cosh\sqrt{\frac{\nu}{4\hat{D}}}},
\end{equation}
\begin{equation}\label{4.12b}
j_{\text{av}}(x)=-\hat{D}y'_{\text{av}}(x)=-T^{\beta+1}\sqrt{\nu \hat{D}}\, \frac{\sinh\left(x\sqrt{\frac{\nu}{\hat{D}}}\right)}
{\cosh\sqrt{\frac{\nu}{4\hat{D}}}}\, .
\end{equation}
\end{subequations}
The average density is readily obtained by combining (\ref{4.8}) and (\ref{4.12}),
\begin{equation}\label{4.13}
    \rho_{\text{av}}(x)=T \left[
    \frac{\cosh\left(x\sqrt{\frac{\nu}{\hat{D}}}\right)}
    {\cosh\sqrt{\frac{\nu}{4\hat{D}}}}\right]^{\frac{1}{\beta+1}}.
\end{equation}
The average dissipation field is basically $y_{\text{av}}(x)$, since
\begin{equation}\label{4.14}
    d_{\text{av}}(x)=-\nu y_{\text{av}}(x).
\end{equation}

\begin{figure}
\centerline{
\includegraphics[width=9cm]{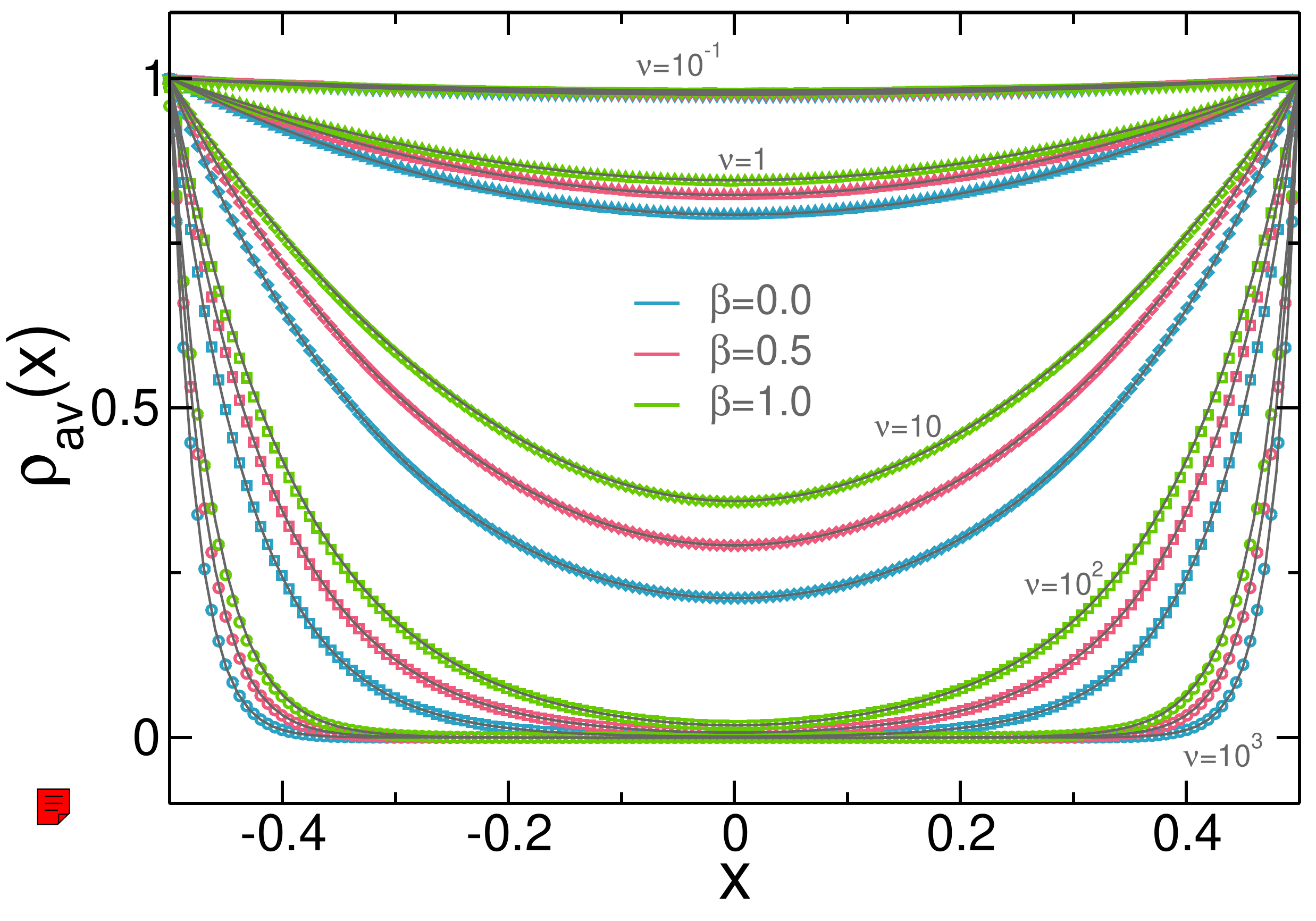}
}
\caption{\small Average density profiles for different values of $\nu\in [10^{-1},10^3]$ and $\beta=0,~0.5,~1$, for $N=160$. The bath temperature is $T=1$. Lines correspond to theoretical predictions, and the agreement is excellent in all cases.}
\label{rho}
\end{figure}

Equation (\ref{4.13}) reveals clearly one of the main properties of dissipative media: the gradients are controlled by the dissipation and not by the boundary conditions. Although both ends of the system are in contact with two heat reservoirs at the same temperature $T$, the density is not constant throughout the system, as it is clearly shown in Fig. \ref{rho}: the higher the macroscopic dissipation coefficient $\nu$ gets, the larger the spatial gradient is.  In the figure, the points correspond to Monte Carlo simulations of the stochastic model introduced in Sec. \ref{s2}, while the lines are the analytical prediction of the hydrodynamic theory, Eq. (\ref{4.13}). The agreement between theory and simulation is excellent in all cases. In the quasielastic limit $\nu\to 0$, the density is almost constant, as $\rho'(x)\propto\nu$ in that limit. This is shown in the figure by the case $\nu=10^{-1}$, in which a weak spatial structure is observed in the scale of the figure, for all values of $\beta$. As $\nu$ increases, the system departs from the ``quasi-homogeneous'' behavior. In fact, for the highest value $\nu=10^3$, there are two boundary layers near the system walls, characterized by a very high internal gradient, while $\rho\to 0$ in the bulk of the system.

\begin{figure}
\centerline{
\includegraphics[width=9cm]{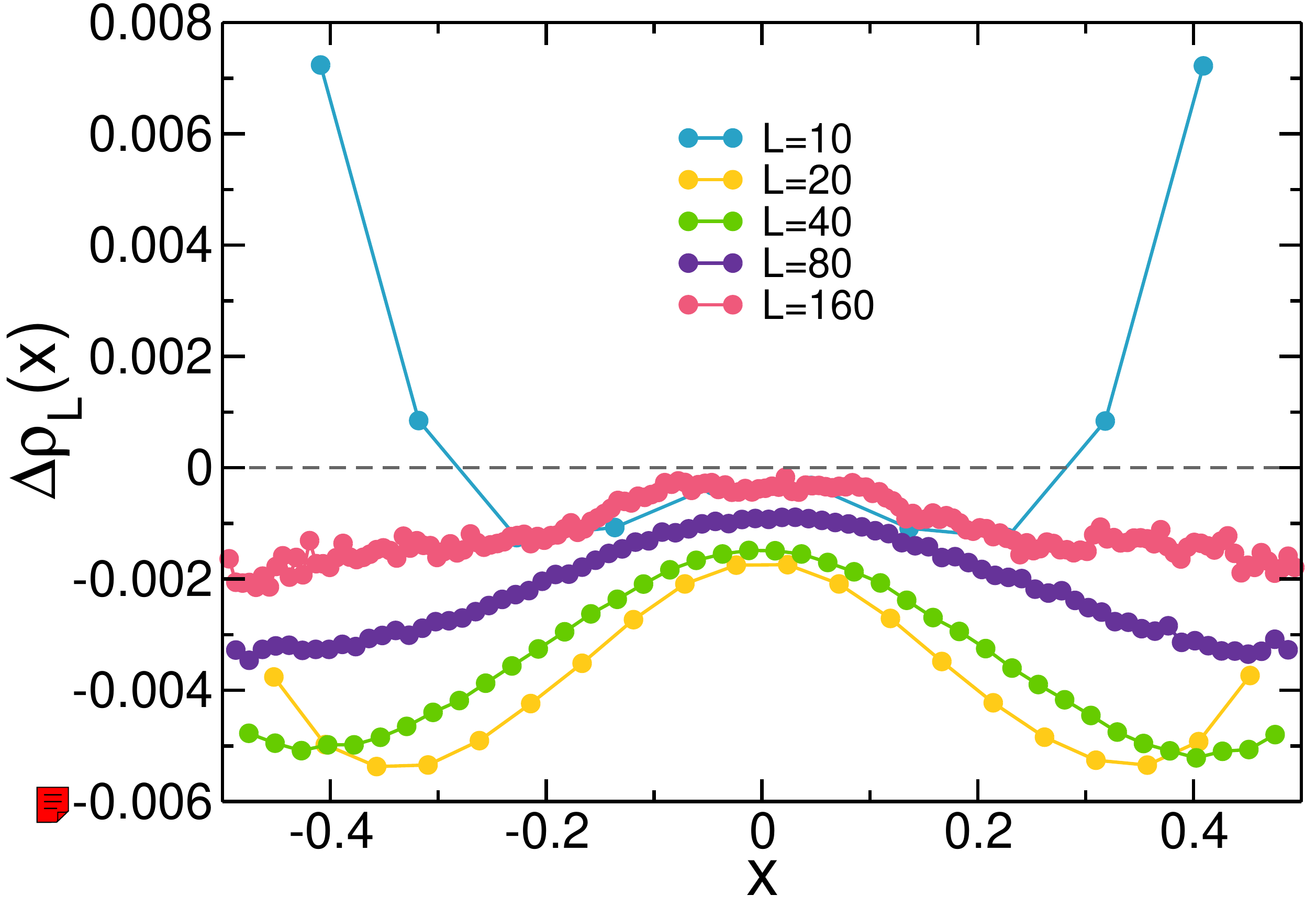}
}
\caption{\small Average excess density profiles for different values of $L\in[10,160]$, $\nu=10$ and $\beta=0.5$. Finite size corrections quickly decay to zero with the system size.}
\label{rhoexcess}
\end{figure}

An important question is the rate of convergence to the hydrodynamic description as the system size is increased. In order give an estimation thereof, we plot in Fig. \ref{rhoexcess} the difference $\Delta \rho_L(x)\equiv \rho_{\textrm{exp}}^{(L)}(x)-\rho_{\textrm{av}}(x)$ between the numerical values measured for the density and the theoretical ones for different system sizes in the range $10\leq L\leq160$, as a function of the spatial coordinate $x$. For the sake of concreteness, we show the data for $\nu=10$ and $\beta=0.5$, for which the spatial gradient is quite important in the system, as seen in Fig. \ref{rho}. $\Delta \rho_L(x)$ exhibits a nontrivial structure which however rapidly decreases with the system size, being very small across the whole system for $L=160$. In fact, even for the smallest size $L=10$, $\Delta \rho_L$ is quite small, since the relative error in under one per cent for all $x$. This means that the hydrodynamic description developed here is valid already for quite small system sizes, although the concrete threshold size depends on the macroscopic dissipation coefficient $\nu$. As $\nu$ increases, larger system sizes must be considered, since the gradients become more important.

\begin{figure}
\centerline{
\includegraphics[width=9cm]{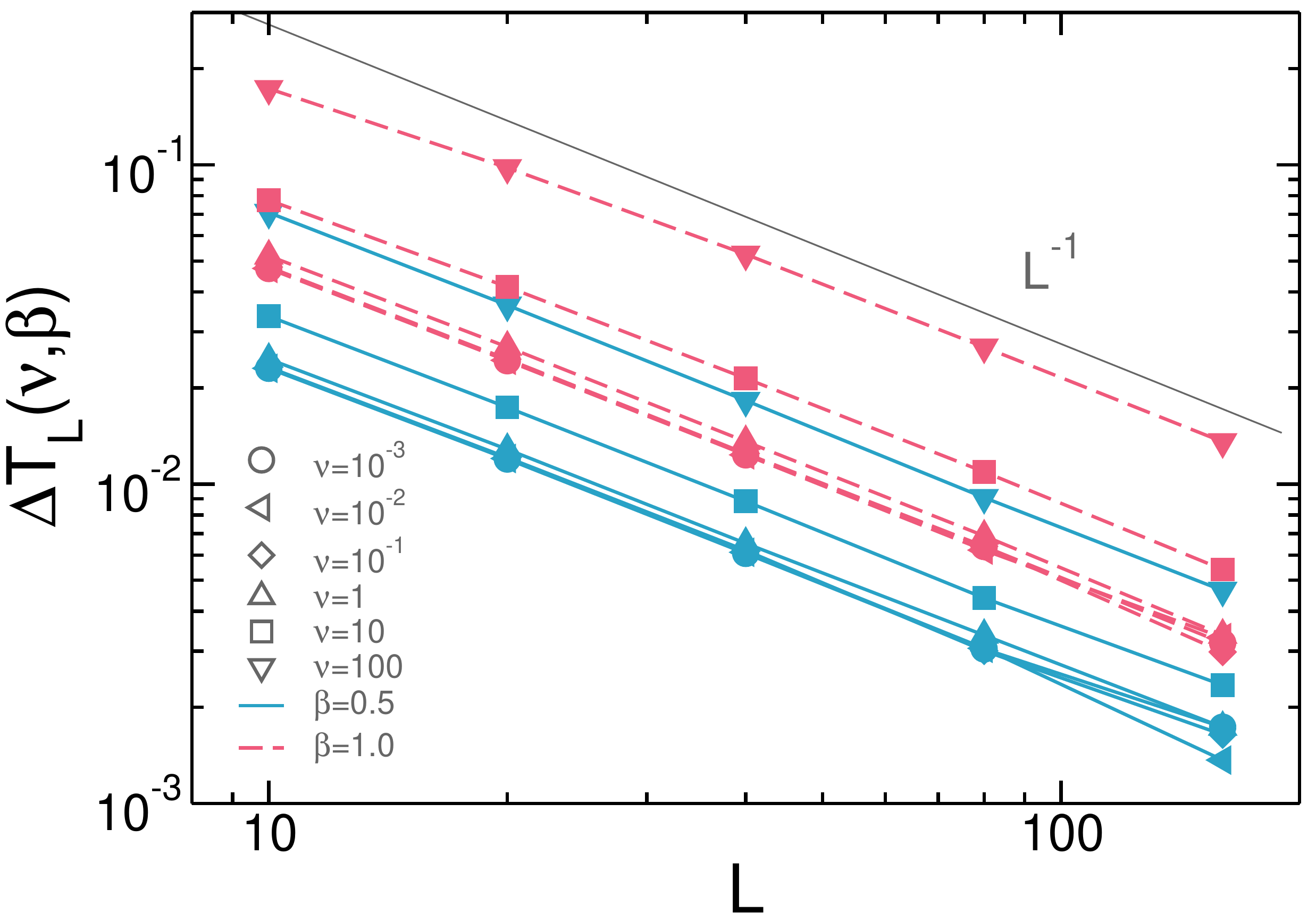}
}
\caption{\small Boundary thermal resistance as a function of the system size $\mathbf{L}$ for different values of $\nu\in[10^{-3},10^2]$ and $\beta=0.5,~1$. In all cases the boundary thermal gap decays as $L^{-1}$, with an amplitude which increases with $\nu$.}
\label{dT}
\end{figure}

A main feature of the nonlinear driven dissipative models studied here is the presence of a nontrivial boundary thermal resistance which shows up as a gap between the average energy of boundary sites and the temperature of the corresponding heat baths, $\Delta T_L\equiv \rho_{\textrm{exp}}^{(L)}(x=\pm\frac{1}{2})-T\neq 0$, see Figs. \ref{rho} and \ref{rhoexcess}. Interestingly, this boundary thermal resistance appears only for $\beta\neq 0$, meaning that the nonlinearity is essential to develop a boundary thermal gap. This phenomenon is well-known to appear in nonlinear energy transport as a result of the scattering of energy carriers when crossing the bath interface, with examples ranging from Fermi-Pasta-Ulam oscillator chains to hard-disks fluids, etc.~\cite{fourier}. Fig. \ref{dT} plots the measured boundary thermal gap as a function of the system size for different values of $\nu$ and $\beta>0$, showing that $\Delta T_L$ decays as $L^{-1}$ for large enough systems, with an amplitude that increases with the dissipation parameter $\nu$ and the nonlinearity $\beta$.

\begin{figure}
\centerline{
\includegraphics[width=9cm]{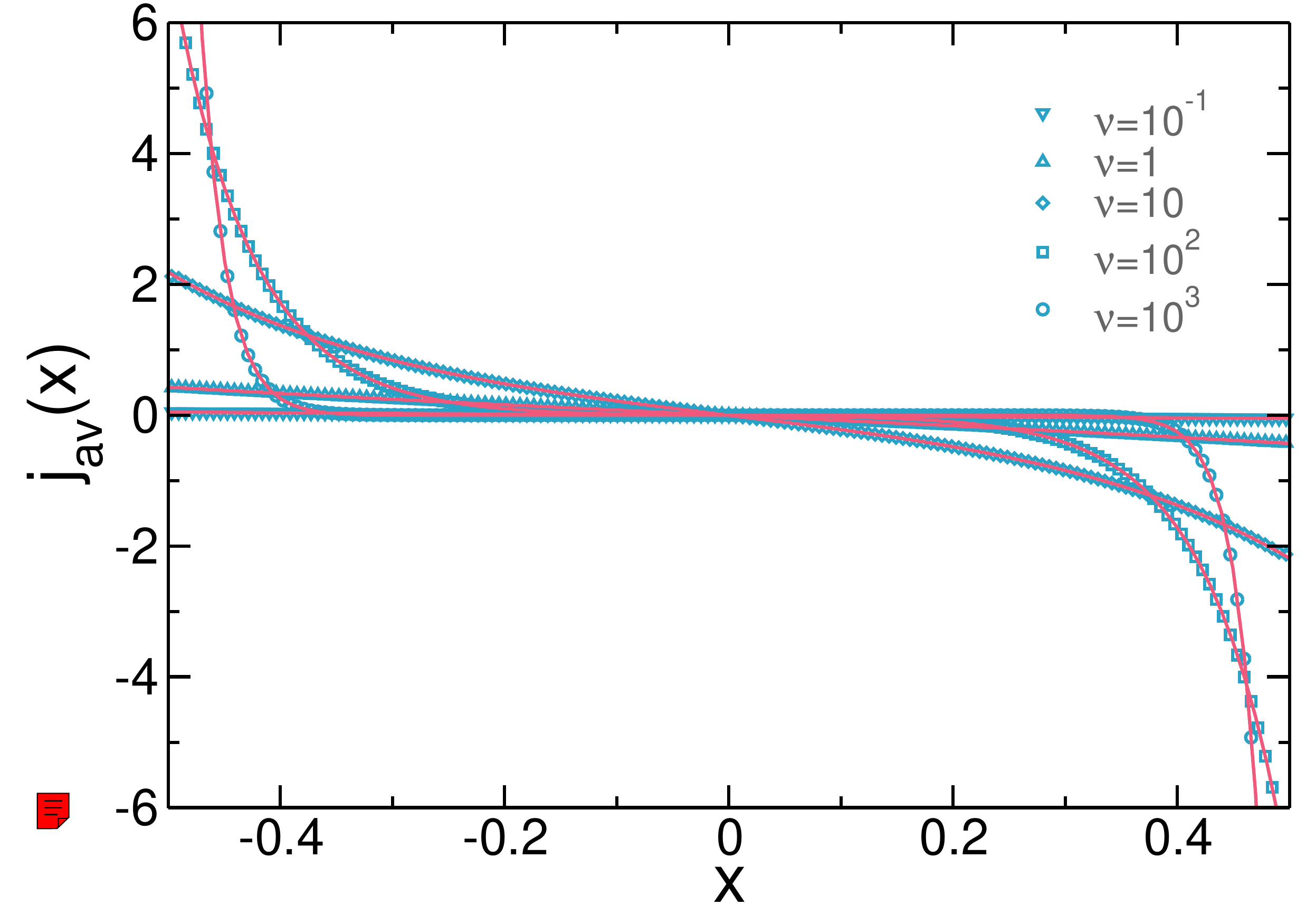}
}
\caption{\small Average current profiles for different values of $\nu\in [10^{-1},10^3]$ and $\beta=0$ for $N=160$. Notice the nontrivial structure due to bulk dissipation (in contrast to conservative systems where the current field is constant across space). Lines correspond to theoretical predictions, and the agreement is excellent in all cases.}
\label{curr}
\end{figure}

Further insight into the hydrodynamic behavior discussed above is given by Fig. \ref{curr}. The average current profile, as measured in simulations after each collision event, is shown for different values of the macroscopic dissipation coefficient $\nu$. For the sake of concreteness, and to make the figure appearance simpler, we have considered $\beta=0$, though other values of $\beta$ exhibit qualitatively similar behavior. The current is not constant throughout the system, in contrast to the situation found in conservative systems. For high dissipation, $\nu\gg 1$, the average current vanishes except in the boundary layers, inside which it is in fact very large. Again, the agreement between the theoretical expression for the current (solid line), Eq. (\ref{4.12b}), and simulation (points) is excellent {in all cases.

\begin{figure}
\centerline{
\includegraphics[width=9cm]{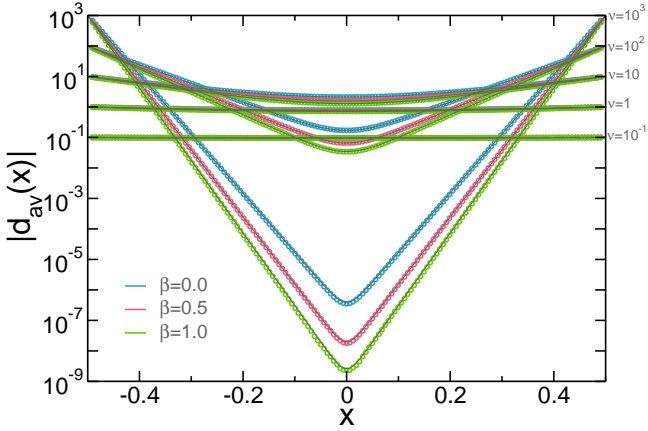}
}
\caption{\small Average dissipation profiles for different values of $\nu\in [10^{-1},10^3]$ and $\beta=0,~0.5,~1$, for $N=160$. The bath temperature is $T=1$. Lines correspond to theoretical predictions, and the agreement is excellent in all cases.}
\label{disip}
\end{figure}

The spatial dependence of the average dissipation field $d_{\text{av}}(x)$ is presented in Fig. \ref{disip}. Note the logarithmic scale in the vertical axis, in order to show more clearly the behavior in the bulk of the system in the  high dissipation regime $\nu\gg 1$. Again the agreement between theory (solid line) and simulation (points) is excellent in all cases, up to the very center of the system, in which $|d_{\text{av}}(x)|$ is really very small when $\nu\gg 1$.

\begin{figure}
\centerline{
\includegraphics[width=9cm]{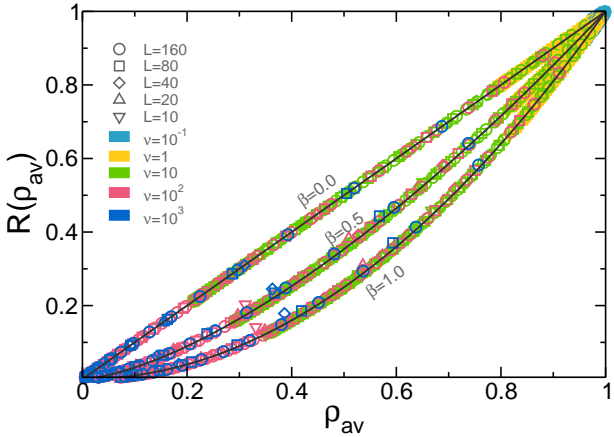}
}
\caption{\small Dissipation transport coefficient $R(\rho)$ measured in simulations as a function of the local average density, for different values of $\nu$, $\beta$ and $L$.  For each value of $\beta$, the numerical data collapse on the theoretical prediction $\rho^{\beta+1}$ $\forall \nu,L$.}
\label{disiprho}
\end{figure}

The simplicity of the model at hand allows us to measure directly in simulations the transport coefficients entering the fluctuating hydrodynamic theory. In particular, a strong implication of this hydrodynamic description is that the transport coefficients are functions of the energy density $\rho$ alone, for large system sizes and under the local equilibrium hypothesis, as expressed by Eqs. (\ref{2.20}), (\ref{2.22}) and (\ref{3.15}). We have tested the validity of this picture by comparing the numerical values of the transport coefficients, measured in Monte Carlo simulations, and the theoretical predictions for the general class of models considered in this section, Eqs. (\ref{4.1}) and (\ref{4.2}). Let us start by considering the numerical evaluation of $R(\rho)$, the new transport coefficient associated to the dissipation. We have built Fig. \ref{disiprho} by plotting the local dissipation along the chain versus the local energy density, i.e. by combining the points of Figs. \ref{rho} and \ref{disip}, and plotting $|d_{\text{av}}(x)|/\nu$ vs $\rho_{\text{av}}(x)$ $\forall x\in[-\frac{1}{2},\frac{1}{2}]$. Aside from the system size $L=160$ considered in Figs. \ref{rho} and \ref{disip}, we have included points obtained for smaller system sizes, ranging from $L=10$ to $L=80$, in order to see the convergence to the hydrodynamic behavior as the system size grows. For each value of $\beta$, all the points collapse onto a master curve, which agrees perfectly with the theoretical expression (\ref{4.1}). This is clear signature of the rapid convergence to the hydrodynamic description with the system size, $L=10$ being enough to recover the theoretical prediction.

\begin{figure}
\centerline{
\includegraphics[width=9cm]{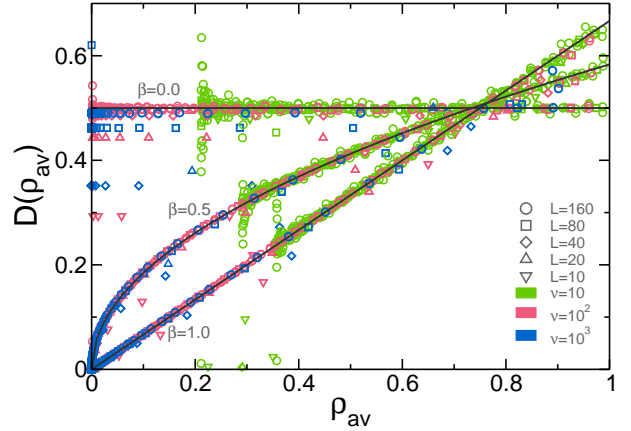}
}
\caption{\small Comparison between the numerical and theoretical values of the diffusivity $D(\rho)$ for $\beta=0,~0.5,~1$. Different values of $\nu$ and $L$ have been considered. The agreement between theory and simulation is excellent. }
\label{diffrho}
\end{figure}

In order to measure numerically the diffusivity, which is just the factor relating the local energy current with the local density gradient according to Fourier's law, we have to combine the data used in Figs. \ref{rho} and \ref{curr} to plot $-j_{\text{av}}/\rho'_{\text{av}}$ (where the prime indicates the numerical spatial derivative of the measured energy profile) vs the average energy density $\rho_{\text{av}}$. As for the dissipation transport coefficient $R(\rho)$, the agreement between simulation and theory is again very good for $D(\rho)$. Notice however that there are some points which depart from the the theoretical prediction, but this departure is due to the smallness of both $j_{\text{av}}(x)$ and $\rho'_{\text{av}}(x)$ near the center of the system, which gives rise to large numerical errors for the quotient $-j_{\text{av}}/\rho'_{\text{av}}$.

\begin{figure}
\centerline{
\includegraphics[width=9cm]{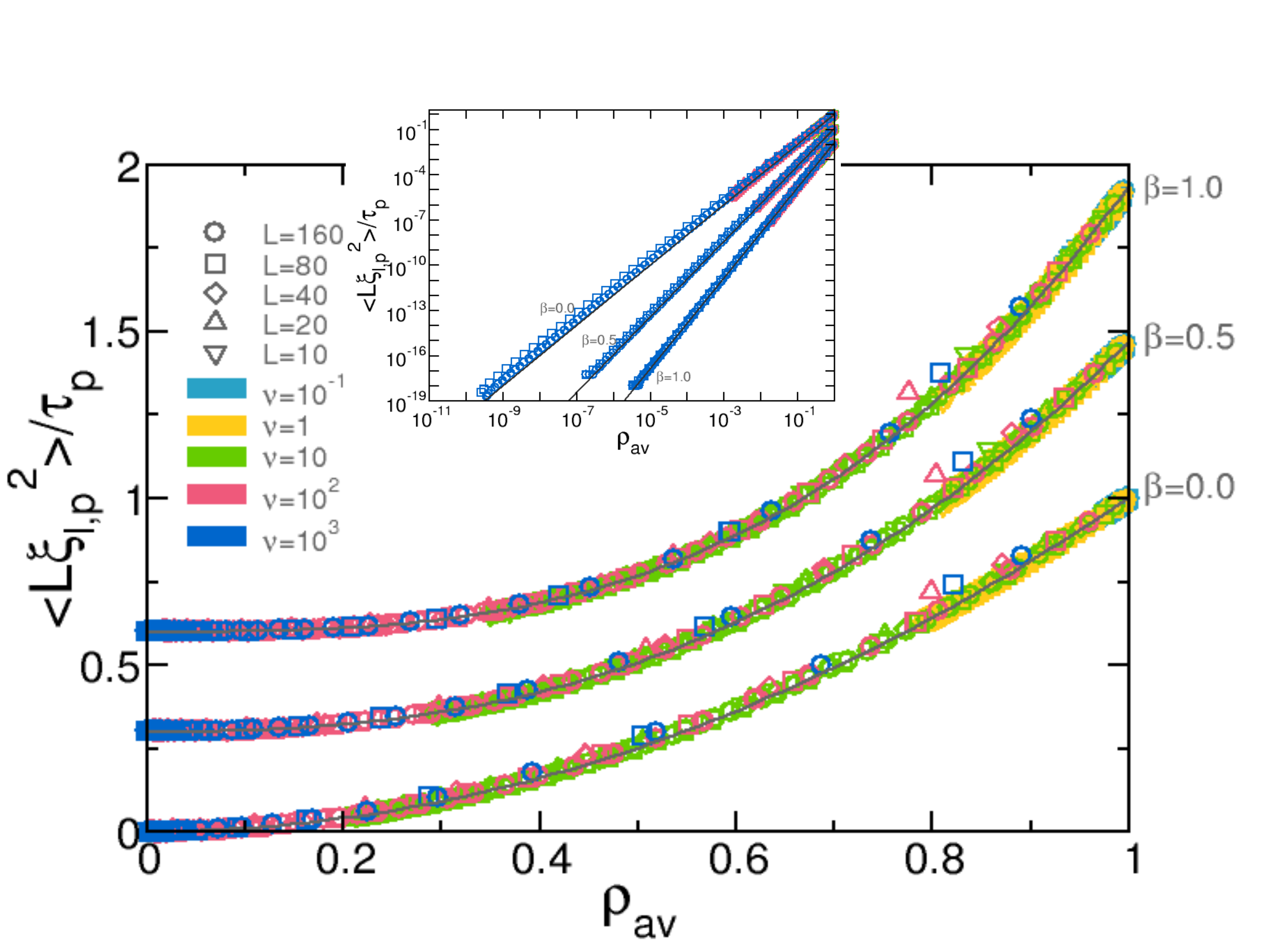}
}
\caption{\small Comparison between the numerical measurement of the mobility $\sigma(\rho)$, as given by Eq. (\ref{4.15}), and the theoretical prediction (\ref{4.3}), for $\beta=0,~0.5,~1$. The agreement is excellent in all cases, even for rather small system sizes ($L=10$). Curves for $\beta=0.5$ ($\beta=1$) have been shifted vertically by $0.3$ ($0.6$) units. The inset shows the equivalent plot in logarithmic scale.}
\label{sigmarho}
\end{figure}

Finally, we have also checked the theoretical prediction for the noise amplitudes, by comparing them to the numerical data obtained in Monte Carlo simulations. Figure \ref{sigmarho} shows the measured amplitude of the current fluctuations, the so-called mobility  $\sigma$. Combination of Eqs. (\ref{3.8}), (\ref{3.9}) and (\ref{3.15}) gives that
\begin{equation}\label{4.15}
  \frac{L \langle \xi_{l,p}^2 \rangle}{\tau_p} = \sigma (\rho_{\text{av}}).
\end{equation}
Therefore, we have plotted the lhs of the above equation as a function of the average density along the chain, both of them numerically measured in simulations. This have been done for the same values of $\beta$ of the previous figures, and for various values of $\nu$ and $L$.  In the inset, the same plot is shown in logarithmic scales. Straight lines with slope $\beta+2$ are observed, in agreement with the algebraic behavior of $\sigma\propto \rho^{\beta+2}$ predicted by the hydrodynamic theory, Eq. (\ref{4.3}). It should be stressed that the excellent agreement between theory and simulation, and particularly the collapse of the data corresponding to different values of $L$,  is a strong proof that the predicted scaling of the current noise with the system size, $\langle \xi(x,t)\xi(x',t')\rangle\propto L^{-1}$ as given by Eq. (\ref{3.14}),  is correct.

Similarly, Fig. \ref{kapparho} shows the measured \emph{diagonal} fluctuations of the dissipation field as a function of the local average density. Following Eqs. (\ref{3.21})-(\ref{3.22}) and the ensuing discussion, these fluctuations are defined as
\begin{equation}
  \frac{L^5 \langle \eta_{l,p}^2 \rangle}{\nu^2 \tau_p} \equiv \kappa_{\textrm{diag}} (\rho_{\text{av}}) = \sigma(\rho_{\text{av}}).
\label{4.15b}
\end{equation}
Again the agreement between theory and simulation results is very good, though finite-size effects, particularly at high densities, are more evident here that for the current fluctuations, see Fig. \ref{sigmarho}. This is natural since dissipation fluctuations are very small and have to be scaled as $L^5$ in the continuum limit, which strongly amplifies the small finite-size effects present in measurements. In any case, Fig. \ref{kapparho} confirms the scaling of the dissipation fluctuations, i.e. that $\la \eta(x,t)\eta(x',t')\ra \propto L^{-3}$ in the continuum limit, as given by Eq. (\ref{3.22}). This supports our claim that dissipation fluctuations are negligible in the $L\to\infty$ limit, where the only source of fluctuations concerns the current field.

\begin{figure}
\centerline{
\includegraphics[width=9cm]{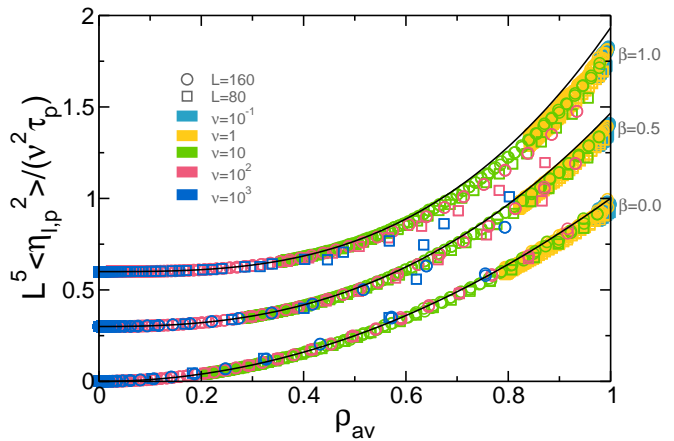}}
\caption{\small Comparison between the numerical measurement of the diagonal part of $\kappa(\rho)$, as given by Eq. (\ref{4.15b}), and the theoretical prediction Eq. (\ref{4.3}), for $\beta=0,~0.5,~1$ and different values of $\nu$ and $L$. Curves for $\beta=0.5$ ($\beta=1$) have been shifted vertically by $0.3$ ($0.6$) units for clarity.
}
\label{kapparho}
\end{figure}

\section{Summary and conclusions}
\label{s5}

In this paper we have studied in detail the fluctuating hydrodynamic theory for a large class of diffusive lattice models characterized by the presence of bulk dissipation and boundary driving. The models dynamics is stochastic: pairs of neighboring lattice sites collide at a rate which is a certain function of the pair energy. A given part of the pair energy is dissipated to the environment, and the remainder thereof is randomly redistributed within the pair. In addition, boundary sites may be coupled to thermal reservoirs which on average inject energy into the system so eventually it reaches a (nonequilibrium) steady state. Thus, this class of driven dissipative models is a generalization of the KMP model for heat conduction to the nonlinear dissipative case.

In the large system size limit, the energy density $\rho(x,t)$ has been shown to obey a mesoscopic fluctuating balance equation, which contains both fluctuating current and dissipation fields, $j(x,t)$ and $d(x,t)$, respectively. The average current field obeys locally a non-linear Fourier's law, with a diffusion coefficient which depends on the density, while the average dissipation becomes a certain function of the local density. Moreover, the current noise is shown to be white and gaussian, while the dissipation noise is subdominant because of the quasi-elasticity of the microscopic dynamics. Detailed expressions for the transport coefficients are derived in this context, using a local equilibrium approximation whose validity has been tested \emph{a posteriori}. Interestingly, the amplitude $\sigma(\rho)$ of the current fluctuations (or mobility) is related to the diffusivity $D(\rho)$ by an Einstein fluctuation-dissipation relation, $\sigma(\rho)=2\rho^2D(\rho)$, despite the fully nonequilibrium character of the problem at hand. Besides, the new transport coefficient associated to the dissipation, $R(\rho)$, is not independent of the diffusion coefficient, but it is related thereto by a first order differential equation.

In order to test in detail the emerging picture, we have applied the general fluctuating hydrodynamic framework developed in the first part of the paper to a particular family of models, in which the collision rate depends algebraically on the energy of the chosen pair. The reasons for this choice are twofold: (i) such power-law dependence mimics the physics of many realistic systems (e.g. for hard-disk fluids the collision rate is proportional to the square root of the local energy density), and (ii) with this choice, the transport coefficients can be explicitly calculated as simple, also algebraic, functions of the energy density. We have checked the predictions of the fluctuating hydrodynamic theory against extensive Monte Carlo simulations of the proposed family of models, and an excellent agreement is found, even for quite small system sizes $L\gtrsim 10$. This agreement strongly supports the validity of the fluctuating hydrodynamic picture and, given the generality of the proposed family of models, a similar situation should be expected for any non-pathological choice of the collision rate function.

The fluctuating hydrodynamic equation here derived is a first step in the complete description of the macroscopic behavior for the broad class of nonequilibrium models introduced in this paper. The governing evolution law, summarized in just three transport coefficients, can now be used in conjunction with the recently-introduced Macroscopic Fluctuation Theory (MFT) \cite{Bertini} to derive the large-deviation functions (LDFs) which control the statistics of fluctuations of the relevant macroscopic observables, in this case the current and the dissipation. LDFs play in nonequilibrium systems a role akin to the free energy or entropy function in equilibrium systems, and hence are of utmost importance to the development of nonequilibrium statistical physics \cite{Bertini}. In a forthcoming paper \cite{PLyH12a} (see also \cite{PLyH11a}) we will use the tools of MFT and advanced Monte Carlo simulations to study in detail the LDF of the relevant macroscopic observables for the general class of nonlinear driven dissipative systems here introduced, thus aiming at a full characterization of macroscopic behavior for this broad family of systems.


\vspace{1cm}

\appendix

\section{Local equilibrium approximation}\label{apa}

In the hydrodynamic fluctuating theory developed in this paper, one of the main goals in to write down a closed fluctuating balance equation for the energy density at the mesoscopic level. At some points in the development of this theoretical framework, we have found that the average current, Eq. (\ref{2.13}), the average dissipation, Eq. (\ref{2.14}), or the amplitude of the current noise, Eq. (\ref{3.9}), are given in terms of certain average values with the general form
\begin{equation}\label{a1}
\left\la \frac{g(\rho_{l,p},\rho_{l+1,p})f(\Sigma_{l,p})}{\Omega_p(L)}\right\ra
\end{equation}
where $g(\rho_{l,p},\rho_{l+1,p})$ are different functions of the energies $\{\rho_{l,p},\rho_{l+1,p}\}$, while $\Sigma_{l,p}$ and $\Omega_p(L)$ have been defined in Eq. (\ref{2.0}) and (\ref{2.b}), respectively. Therefore,
\begin{widetext}
\begin{eqnarray}
    \left\la \frac{g(\rho_{l,p},\rho_{l+1,p})f(\rho_{l,p}+\rho_{l+1,p})}{\Omega_p(L)}
    \right\ra&=& \int_0^\infty \!\! d\rho_l \int_0^\infty \!\! d\rho_{l+1}\,  g(\rho_{l},\rho_{l+1})f(\rho_{l}+\rho_{l+1}) P(\rho_l,\rho_{l+1};p) \nonumber\\
     \times   \int_0^\infty \!\! d\rho_1 &\cdots &  \int_0^\infty \!\! d\rho_{l-1} \int_0^\infty \!\! d\rho_{l+2}\cdots \int_0^\infty \!\! d\rho_N \, \Omega^{-1}(L) P(\rho_1,\ldots,\rho_{l-1},\rho_{l+2},\ldots,\rho_N|\rho_l,\rho_{l+1};p).
     \nonumber\\
     &&\label{a3}
\end{eqnarray}
\end{widetext}
In order to write the equation above, we have used Bayes theorem to write the probability that the system is in configuration $\bm{\rho}=\{\rho_1,\rho_2,\ldots,\rho_N\}$ at time step $p$ as $P(\rho_l,\rho_{l+1};p)P(\rho_1,\ldots,\rho_{l-1},\rho_{l+2},\ldots,\rho_N|\rho_l,\rho_{l+1};p)$;  in which $P(\rho_l,\rho_{l+1};p)$ is the probability of finding sites $l$ and $l+1$ with energies $\rho_l$ and $\rho_{l+1}$ at time $p$, independently of the configuration of the remainder of the sites, and $P(\rho_1,\ldots,\rho_{l-1},\rho_{l+2},\ldots,\rho_N|\rho_l,\rho_{l+1};p)$ is the conditional probability of finding the remainder of the sites in configuration $\{\rho_1,\ldots,\rho_{l-1},\rho_{l+2}\ldots,\rho_N\}$ at time $p$, provided that sites $l$ and $l+1$ have energies $\rho_l$ and $\rho_{l+1}$, respectively.

First, let us calculate the integral over $\{\rho_1,\ldots,\rho_{l-1},\rho_{l+2}\ldots,\rho_N\}$ in Eq. (\ref{a3}). From the definition of $\Omega_p(L)$, Eq. (\ref{2.b}),
\begin{widetext}
\begin{equation}
    \Omega(L)=\frac{1}{L}\left[\sum_{l'=1}^{l-2} f(\rho_{l'}+\rho_{l'+1})+f(\rho_{l-1}+\rho_{l+2})+\sum_{l'=l+2}^L f(\rho_{l'}+\rho_{l'+1})\right]+\frac{1}{L}\left[\sum_{l'=l-1}^{l+1} f(\rho_{l'}+\rho_{l'+1})-f(\rho_{l-1}+\rho_{l+2})\right].
    \label{a4}
\end{equation}
\end{widetext}
Note that, in Eq. (\ref{a4}),  we write $\Omega(L)$ and $\rho_i$, omitting the subindex corresponding to the time step $p$. This is so because we will insert Eq. (\ref{a4}) into (\ref{a3}), where the integration variables are dummy, $\{\rho_1,\rho_2,\ldots,\rho_N\}$, and the time dependence appears in the probability distributions. In fact, we have already used this notation in Eq. (\ref{a3}), for $\Omega_p(L)$ but also for the functions $g$ and $f$. Equation (\ref{a4}) implies that
\begin{widetext}
\begin{equation}
  \Omega(L) = \frac{L-2}{L} \Omega^*(L-2)+\frac{1}{L}\left[\sum_{l'=l-1}^{l+1} f(\rho_{l'}+\rho_{l'+1})-f(\rho_{l-1}+\rho_{l+2})\right],
   \label{a5}
\end{equation}
where $\Omega^*(L-2)$ corresponds to a system with size $L-2$, without the sites $l$ and $l+1$. Equation (\ref{a5}) suggests that, in the large system size limit,
\begin{equation}\label{a6}
  \int_0^\infty d\rho_1 \cdots \int_0^\infty d\rho_{l-1} \int_0^\infty d\rho_{l+2}\cdots\int_0^\infty d\rho_N \, \Omega^{-1}(L) P(\rho_1,\ldots,\rho_{l-1},\rho_{l+2},\ldots,\rho_N|\rho_l,\rho_{l+1};p)
    \sim \la {\Omega_p^*}^{-1}(L-2)\ra \sim \tau_p,
\end{equation}
\end{widetext}
where $\tau_p=\la \Omega_p^{-1}(L)\ra$ for $L\to\infty$, as defined in Eq. (\ref{2.16b}). The subindex $p$ reappears in $\tau_p$, because the average value is calculated at time $p$, with the probability distribution $P(\rho_1,\rho_2,\ldots,\rho_N;p)$. Thus, substituting Eq. (\ref{a6}) into (\ref{a3}),
\begin{widetext}
\begin{equation}\label{a7}
    \left\la \frac{g(\rho_{l,p},\rho_{l+1,p})f(\rho_{l,p}+\rho_{l+1,p})}{\Omega_p(L)}\right\ra\sim \tau_p \left\la g(\rho_{l,p},\rho_{l+1,p})f(\rho_{l,p}+\rho_{l+1,p}) \right\ra.
\end{equation}
This result will be repeatedly used in the following sections to simplify the calculations.
In order to further advance, we have to calculate the average
\begin{equation}\label{a8}
    \left\la g(\rho_{l,p},\rho_{l+1,p})f(\rho_{l,p}+\rho_{l+1,p}) \right\ra=\int_0^\infty d\rho_l \int_0^\infty d\rho_{l+1}\,  g(\rho_{l},\rho_{l+1})f(\rho_{l}+\rho_{l+1}) P(\rho_l,\rho_{l+1};p),
\end{equation}
for which the distribution function $P(\rho_l,\rho_{l+1};p)$ must be known. At this point, and so as to calculate these averages, we introduce the \textit{local equilibrium approximation}, i.e. we assume that the probability distribution $P(\rho_l,\rho_{l+1};p)$ inside the integral in Eq. (\ref{a8}) can be substituted by
\end{widetext}
\begin{eqnarray}
    P_{\text{LE}}(\rho_l,\rho_{l+1};p)&=&\frac{1}{\la \rho_{l,p} \ra\la \rho_{l+1,p}\ra} \nonumber \\
    && \times \exp\left(-\frac{\rho_{l}}{\la \rho_{l,p} \ra}-\frac{\rho_{l+1}}{\la \rho_{l+1,p}\ra}\right).\label{a9}
\end{eqnarray}
This local equilibrium hypothesis allows us to calculate expressions of the averages in terms of local energy densities $\la\rho_{l,p}\ra$ and $\la\rho_{l+1,p}\ra$. Furthermore, it will be assumed (when necessary) that $\la\rho_{l+1,p}-\rho_{l,p}\ra={\cal O}(L^{-1})$, consistently with the spatial continuum limit introduced in sec. \ref{s2}. In this way, the average in Eq. (\ref{a8}) becomes a functional of the energy density field $\rho_{\text{av}}(x,t)$,
\begin{eqnarray}
\la\rho_{l+1,p}\ra & = & \la \rho_{l,p}\ra+L^{-1} \frac{\la \rho_{l+1,p}-\rho_{l,p}\ra}{\Delta x} \nonumber\\
&\sim & \rho_{\text{av}}(x,t)+L^{-1}\partial_x \rho_{\text{av}}(x,t).\label{a10}
\end{eqnarray}
since
\begin{equation}\label{a10b}
    \frac{\la \rho_{l+1,p}-\rho_{l,p}\ra}{\Delta x}\sim \partial_x \rho_{\text{av}}(x,t)
\end{equation}
in the spatial and time continuum limits we are considering throughout the paper.

\subsection{Derivation of equation (\ref{2.19})} 

Our starting point is Eq. (\ref{2.13}) for the current,
\begin{eqnarray}
    \la j_{l,p}\ra&= & \frac{\alpha}{2L} \left\la  \frac{(\rho_{l,p}-\rho_{l+1,p}) f(\Sigma_{l,p})}{\Omega_p(L)} \right\ra \nonumber\\
     &\sim &\frac{\alpha}{2L} \tau_p \la (\rho_{l,p}-\rho_{l+1,p}) f(\Sigma_{l,p}) \ra, \label{a11}
\end{eqnarray}
where we have made use of Eq. (\ref{a7}). Therefore, we have to evaluate the average $\la (\rho_{l,p}-\rho_{l+1,p}) f(\Sigma_{l,p}) \ra$ in the local equilibrium approximation. Up to first order in $L^{-1}$,
\begin{widetext}
\begin{equation}
    \la (\rho_{l,p}-\rho_{l+1,p}) f(\Sigma_{l,p}) \ra_{\text{LE}}\sim  - L^{-1} \frac{\partial_x \rho_{\text{av}}}{\rho_{\text{av}}^3} \int_0^\infty d\rho_l \int_0^\infty d\rho_{l+1} (\rho_l-\rho_{l+1}) \left(1-\frac{\rho_{l+1}}{\rho_{\text{av}}}\right) f(\rho_l+\rho_{l+1})
 \exp\left(-\frac{\rho_l+\rho_{l+1}}{\rho_{\text{av}}}\right),
    \label{a13}
\end{equation}
where we have taken into account Eqs.(\ref{a9}) and (\ref{a10}). By interchanging the dummy integration variables $\rho_l$ and $\rho_{l+1}$, an expression symmetric to the previous one is obtained. Averaging the two equivalent expressions, the above equation is transformed into
\begin{eqnarray}
    \la (\rho_{l,p}-\rho_{l+1,p}) f(\Sigma_{l,p}) \ra_{\text{LE}} &\sim & -L^{-1} \frac{\partial_x \rho_{\text{av}}}{2 \rho_{\text{av}}^4} \int_0^\infty d\rho_l \int_0^\infty d\rho_{l+1} (\rho_l-\rho_{l+1})^2 f(\rho_l+\rho_{l+1})\exp\left(-\frac{\rho_l+\rho_{l+1}}{\rho_{\text{av}}}
    \right)\nonumber \\
    & =& - L^{-1} \frac{\partial_x \rho_{\text{av}}}{2} \int_0^\infty dx \int_0^\infty dy (x-y)^2 f(\rho_{\text{av}}(x+y)) e^{-(x+y)}.    \label{a14}
\end{eqnarray}
\end{widetext}
where we have made use of $\la \rho_{l+1,p}\ra\sim \la \rho_{l,p}\ra\to \rho_{\text{av}}(x,t)$, as given by Eq. (\ref{a10}), in the large system size limit. Besides, in order to obtain the last equality in Eq. (\ref{a14}), we have introduced the change of variables $\rho_l=\rho_{\text{av}}x$, $\rho_{l+1}=\rho_{\text{av}}y$. We can further simplify this expression by going to polar coordinates $(r,\phi)$, and defining $x=a^2$, $y=b^2$, with $a=r\cos\phi$, $b=r\sin\phi$. Thus, the integral over the angle variable $\phi$ can be carried out, with the result
\begin{equation}\label{a15}
    \la (\rho_{l,p}-\rho_{l+1,p}) f(\Sigma_{l,p}) \ra_{\text{LE}}=-\frac{\partial_x \rho_{\text{av}}}{3L} \int_0^\infty \!\! dr r^7 f(\rho_{\text{av}}r^2) e^{-r^2}.
\end{equation}
The equation above, together with Eqs. (\ref{a11}), is equivalent to Eqs. (\ref{2.19})-(\ref{2.20}).

\subsection{Derivation of equation (\ref{2.21})}

Now, we will start from Eq. (\ref{2.14}) for the dissipation,
\begin{eqnarray}
    \la d_{l,p} \ra & = &  \frac{1-\alpha}{L} \left\la \rho_{l,p} \left[ \frac{f(\Sigma_{l,p})+f(\Sigma_{l-1,p})}{\Omega_p(L)}\right] \right\ra \nonumber \\
    &\sim & \frac{1-\alpha}{L}\tau_p \la \rho_{l,p} \left[ f(\Sigma_{l,p})+f(\Sigma_{l-1,p})\right]\ra .\nonumber\\
&&    \label{a16}
\end{eqnarray}
where we have used again Eq. (\ref{a7}). Similarly to what has been done in the previous subsection, we calculate now the average $\left\la \rho_{l,p} \left[ f(\Sigma_{l,p})+f(\Sigma_{l-1,p})\right]\right\ra$ in the local equilibrium approximation. In fact, we have a simpler case, because this average is of the order of unity in the large system size limit. It is easily shown that
\begin{eqnarray}
    && \la \rho_{l,p} \left[ f(\Sigma_{l,p})+f(\Sigma_{l-1,p})\right]\ra_{\text{LE}}=\rho_{\text{av}} \nonumber   \\ && \quad \times \int_0^\infty dx \int_0^\infty dy (x+y) f(\rho_{\text{av}}(x+y)) e^{-(x+y)}.\nonumber\\
    &&  \label{a19}
\end{eqnarray}
As already done when writing Eq. (\ref{a14}), we have taken into account $\la \rho_{l+1,p}\ra\sim \la \rho_{l,p}\ra\to \rho_{\text{av}}(x,t)$, Eq. (\ref{a10}), and introduced the change of variables $\rho_l=\rho_{\text{av}}x$, $\rho_{l+1}=\rho_{\text{av}}y$. Now, by introducing the same change of variables to polar coordinates as in the previous subsection, and making the integral over the angle variable, we arrive at
\begin{eqnarray}
     \la \rho_{l,p} \left[ f(\Sigma_{l,p})+f(\Sigma_{l-1,p})\right]\ra_{\text{LE}}=2\rho_{\text{av}} \nonumber &&  \\ \times \int_0^\infty \!\! dr r^5 f(\rho_{\text{av}} r^2) e^{-r^2}, \label{a20} &&
\end{eqnarray}
that implies Eq. (\ref{2.21}), taking into account Eqs. (\ref{a16}).

\subsection{Derivation of Eq. (\ref{3.10})}

Now, we evaluate the average in Eq. (\ref{3.9}),
\begin{eqnarray}
    \Xi_{l,p} & = & \frac{\alpha^2}{3}\large\la \frac{\left(\rho_{l+1,p}^2+\rho_{l,p}^2-\rho_{l,p}\rho_{l+1,p}\right)
    f(\Sigma_{l,p})}{\Omega_p(L)}
    \large\ra \nonumber \\
    & \sim & \frac{\alpha^2}{3}\tau_p \left\la \left(\rho_{l+1,p}^2+\rho_{l,p}^2-\rho_{l,p}\rho_{l+1,p}\right)
     f(\Sigma_{l,p}) \right\ra \nonumber \\
     && \label{a21}
\end{eqnarray}
along the same lines of the previous subsections. By introducing the local equilibrium approximation,
\begin{eqnarray}
    && \left\la \left(\rho_{l+1,p}^2+\rho_{l,p}^2-\rho_{l,p}\rho_{l+1,p}\right)f(\Sigma_{l,p}) \right\ra_{\text{LE}}=\rho_{\text{av}}^2 \nonumber \\
    && \quad \times\int_0^\infty dx \int_0^\infty dy \, (x^2+y^2-xy) f(\rho_{\text{av}} (x+y)) e^{-(x+y)}, \nonumber \\
    && \label{a24}
\end{eqnarray}
where we have used again the same simplifications as above.
 After going to polar coordinates and evaluating the angle integral, one arrives at
\begin{eqnarray}
    \la\left(\rho_{l+1,p}^2+\rho_{l,p}^2-\rho_{l,p}\rho_{l+1,p}\right)f(\Sigma_{l,p}) \ra_{\text{LE}} =\rho_{\text{av}}^2  \nonumber && \\
    \times \int_0^\infty dr r^7 f(\rho_{\text{av}}r^2) e^{-r^2}.  \label{a25} &&
\end{eqnarray}
Inserting Eq. (\ref{a25}) into Eq. (\ref{a21}) leads to the desired result (\ref{3.10}).

\section{Gaussian character of the current noise}
\label{apb}

In the large system size limit $L\gg 1$, the current noise introduced in section \ref{s3b} is white,
\begin{equation}\label{b1}
    \la \xi(x,t)\ra=0 , \quad \la \xi(x,t)\xi(x',t')\ra =\frac{1}{L} \sigma(\rho) \delta(x-x') \delta(t-t'),
\end{equation}
with the mobility $\sigma(x,t)$ given by Eq. (\ref{3.15}), independent of $L$. We can introduce a new noise field $\widetilde{\xi}(x,t)$ by
\begin{equation}\label{b2}
    \xi(x,t)=L^{-1/2} \widetilde{\xi}(x,t),
\end{equation}
and $\widetilde{\xi}(x,t)$ remains finite in the large system size limit as $L\to\infty$,
\begin{equation}\label{b3}
    \la \widetilde{\xi}(x,t)\ra=0 , \quad \la \widetilde{\xi}(x,t)\widetilde{\xi}(x',t')\ra =\sigma(\rho) \delta(x-x') \delta(t-t').
\end{equation}
In this appendix, we show that all the higher-order cumulants of $\widetilde{\xi}(x,t)$ vanish in the thermodynamic limit as $L\to\infty$. Let us consider a cumulant of order $n$ of the microscopic noise $\xi_{l,p}$, that is equal to the $n$-th order moment of $\xi$ plus a sum of nonlinear products of lower order moments of $\xi$. A calculation analogous to the one carried out for the correlation $\la \xi_{l,p}\xi_{l',p'}\ra$ shows that the leading behavior of any moment is of the order of $L^{-1}$, which is obtained when all the times are the same. Therefore, the moment $\la j_{l,p} j_{l',p'} \cdots j_{l^{(n)},p^{(n)}}\ra$ gives the leading behavior, of the order of $L^{-1}$ for $p=p'=\cdots=p^{(n)}$; any other contribution to the cumulant is at least of the order of $L^{-2}$. Therefore,
\begin{widetext}
\begin{equation}\label{b4}
    \la j_{l,p} j_{l',p'} \cdots j_{l^{(n)},p^{(n)}}\ra \sim L^{-1} \tau_p  \la C_{lp}\ra \delta_{l,l'} \delta_{l',l''}\cdots\delta_{l^{(n-1)},l^{(n)}} \delta_{p,p'} \delta_{p',p''}\cdots\delta_{p^{(n-1)},p^{(n)}},
\end{equation}
where $\la C_{lp}\ra$ is a certain average which remains finite in the large system size limit as $L\to\infty$. In the continuum limit, each current introduces a factor $L^2/\tau_p$ due to the scaling introduced in Eq. (\ref{2.16a}),  and
\begin{eqnarray}
     \la \xi(x,t) \xi(x',t')\cdots \xi(x^{(n)},t^{(n)})\ra & \sim &  L^{2n-1} \tau_p^{1-n} \la C(x,t)\ra  \frac{\delta(x-x')\delta(x'-x'')\cdots\delta(x^{(n-1)}-x^{(n)})}{L^{n-1}}
    \nonumber\\
    &&
     \times
    \frac{\tau_p^{n-1}\delta(t-t')\delta(t'-t'')\cdots\delta(t^{(n-1)}-t^{(n)})}{L^{3(n-1)}} \label{b5a}
\end{eqnarray}
i.e.
\begin{eqnarray}
   \la \xi(x,t) \xi(x',t')\cdots \xi(x^{(n)},t^{(n)})\ra &=&  L^{3-2n}
   \la C(x,t)\ra \delta(x-x')\delta(x'-x'')\cdots\delta(x^{(n-1)}-x^{(n)})
    \nonumber \\  && \times \delta(t-t')\delta(t'-t'')\cdots\delta(t^{(n-1)}-t^{(n)}), \label{b5b}
\end{eqnarray}
where $\la C(x,t)\ra$ is the (finite) leading behavior of $\la C_{l,p}\ra$ in the large system size limit $L\gg 1$, under the local equilibrium approximation introduced in section \ref{s3a}. We have also taken into account the relationship between Kronecker and Dirac deltas, Eq. (\ref{3.16}). Going to the rescaled, of the order of unity, noise $\widetilde{\xi}$,
\begin{eqnarray}
    \la \widetilde{\xi}(x,t) \widetilde{\xi}(x',t')\cdots \widetilde{\xi}(x^{(n)},t^{(n)})\ra &\sim & L^{3\left(1-\frac{n}{2}\right)}
   \la C(x,t)\ra \delta(x-x')\delta(x'-x'')\cdots\delta(x^{(n-1)}-x^{(n)}) \nonumber \\
   && \times \delta(t-t')\delta(t'-t'')\cdots\delta(t^{(n-1)}-t^{(n)}). \label{b6}
\end{eqnarray}
\end{widetext}
Therefore, in the limit as $L\to\infty$,
\begin{equation}\label{b7}
    \la \widetilde{\xi}(x,t) \widetilde{\xi}(x',t')\cdots \widetilde{\xi}(x^{(n)},t^{(n)})\ra =0 \quad \mbox{for all $n>2$},
\end{equation}
which completes the proof. A similar calculation can be carried out for the dissipation noise, that is also gaussian in the thermodynamic limit. Nevertheless, it will not be presented here, since it is subdominant as compared to the current noise for large system sizes $L\gg 1$, and therefore has been neglected.

\begin{acknowledgments}
We acknowledge financial support from Spanish Ministerio de Econom\'{\i}a y Competitividad projects FIS2011-24460 and FIS2009-08451, EU-FEDER funds, and Junta de Andaluc\'{\i}a projects P07-FQM02725 and P09-FQM4682. AP would like to thank Prof. Emmanuel Trizac for helpful discussions during a short stay at Universit\'e Paris-Sud (Orsay) in summer 2011. AL acknowledges Prof. P. L. Garrido for his support during a stay at Dpto.\ de Electromagnetismo and F\'{\i}sica de la Materia of the University of Granada in 2010, when this work was initiated.
\end{acknowledgments}


\begin{thebibliography}{99}
\bibitem{foot1} Note that Eq. (\ref{3.1}) is valid for each realization of the stochastic process. The main term $\widetilde{j}_{l,p}$ can be interpreted as the conditional average of $j_{l,p}$ given the system is in configuration $\rho$. Under this interpretation, it is clear why the averages of $j_{l,p}$ and $\widetilde{j}_{l,p}$ coincide.
\bibitem{foot2} We restrict ourselves to a subset of trajectories which correspond to the same density profile $\rho(x,t)$. The same analysis that has been carried out here can be done, but $\rho_{\text{av}}(x,t)$ in Eq. (\ref{3.14}) should be replaced by $\rho(x,t)$.
\bibitem{gamma} $\Gamma(z)$ is the gamma function, $\Gamma(z)=\int_0^\infty dt \,t^{z-1} e^{-t}$, with the property $\Gamma(z+1)=z\Gamma(z)$
\end{thebibliography}
\end{document}